\documentclass[draftcls,11pt,onecolumn]{IEEEtran}
\usepackage{amsfonts}
\usepackage{amssymb}
\usepackage[cmex10]{amsmath}
\usepackage{graphicx}
\usepackage{subfigure}
\usepackage{verbatim}
\usepackage{cite}
\usepackage{booktabs}
\usepackage[ruled]{algorithm2e}
\usepackage{multirow}
\usepackage{bm}
\usepackage{diagbox}
\usepackage{tabularx,booktabs}
\usepackage{amsthm}
\usepackage{booktabs}
\usepackage{multirow}
\usepackage{setspace}
\usepackage{stfloats}
\usepackage{fancyhdr}
\usepackage{float}
 \usepackage{enumerate}
\begin{document}
\title{Suboptimum Low Complexity Joint Multi-target Detection and Localization for Noncoherent MIMO Radar with Widely Separated Antennas}

\author{Wei~Yi$^*$,~\IEEEmembership{Member,~IEEE,}
        Tao~Zhou,
        Mingchi~Xie,~\IEEEmembership{Member,~IEEE,}
        Yue~Ai
        and~Rick S. Blum,~\IEEEmembership{Fellow,~IEEE}
\thanks{The work of R. S. Blum was supported by the National Science
Foundation under Grant No. ECCS-1405579.}
\thanks{W. Yi, T. Zhou, M. Xie and Y. Ai are with the University of Electronic Science and Technology of China, Chengdu 611731, China. W. Yi is the corresponding author (e-mail: kussoyi@gmail.com).}
\thanks{R. S. Blum is with the Electrical and Computer Engineering Department, Lehigh University, Bethlehem, PA 18015 USA (e-mail: rblum@eecs.lehigh.edu) }}

\markboth{IEEE Transactions on Signal Processing}%
%IEEE Transactions on Aerospace and Electronic Systems
 {Shell\MakeLowercase{\textit{et al.}}: Bare Demo of IEEEtran.cls for
Journals}

\maketitle

\begin{abstract}
In this paper, the problems of simultaneously detecting and localizing multiple targets are considered for noncoherent multiple-input multiple-output (MIMO) radar with widely separated antennas. By assuming a prior knowledge of target number, an optimal solution to this problem is presented first. It is essentially a maximum-likelihood (ML) estimator searching parameters of interest in a high-dimensional space. However, the complexity of this method increases exponentially with the number $G$ of targets. Besides, without the prior information of the number of targets, a multi-hypothesis testing strategy to determine the number of targets is required, which further complicates this method. Therefore, we split the joint maximization into $G$ disjoint optimization problems by clearing the interference from previously declared targets. In this way, we derive two fast and robust suboptimal solutions which allow trading performance for a much lower implementation complexity which is almost independent of the number of targets. In addition, the multi-hypothesis testing is no longer required when target number is unknown. Simulation results show the proposed algorithms can correctly detect and accurately localize multiple targets even when targets share common range bins in some paths.
\end{abstract}

\begin{center} \bfseries EDICS Category: RAS-DMMP, RAS-LCLZ.\end{center}
%\begin{Keywords}
%noncoherent MIMO radar; maximum-likelihood estimation; multi-target detection and localization;
%\end{Keywords}
%\mytableofcontents
%\tableofcontents

\setlength {\abovedisplayskip} {6pt plus 3.0pt minus 4.0pt}
\setlength {\belowdisplayskip} {6pt plus 3.0pt minus 4.0pt}

\section{Introduction}
%Recently, multiple-input multiple-output (MIMO) radar, inspired by wireless communications, has drawn more and more attention from researchers. For MIMO radar system, the waveform diversity and spatial diversity are fully exploited to obtain better radar performance in many aspects than conventional radar system. Due to the different mode of placement of antennas, generally MIMO radar can be classified into two categories. The first category is called co-located MIMO radar with closely spaced antennas, which can obtain waveform diversity via transmitting multiple independent signals \cite{Li}, and its architecture is similar to conventional phase array radar \cite{Lya}. The second category, known as MIMO radar with widely separated antennas, employs widely separated antennas to observe a target at different angles of  view to achieve the spatial diversity \cite{Haimovich}. For MIMO radar with widely separated antennas, there are two different way in signal processing, noncoherent and coherent processing. The noncoherent processing requires time synchronization and no phase synchronization while the coherent processing requires both of them. It has been shown that MIMO radar with widely separated antennas can mitigate the performance degradations caused by the radar cross-section fluctuations \cite{Fishler} \cite{Lehmann}, and can provide both diversity gain and geometry gain to improve detection and localization performance \cite{E.Fishler}. The more detailed and deeper research about the concept and architecture of MIMO radar system are discussed in \cite{LiAndStoica}.

Recently, multiple-input multiple-output (MIMO) radar, inspired by wireless communications, has drawn more and more attention from researchers \cite{Li,Lya,Haimovich,Fishler1,LiAndStoica}. Generally, MIMO radar can be classified into two categories, namely, co-located MIMO radar \cite{Li} and  MIMO radar with widely separated antennas \cite{Haimovich}. The former one, similar to conventional phase array radar \cite{Lya}, employs multiple independent signals transmitted by the closely spaced antennas to obtain waveform diversity \cite{Lehmann}. The latter one, observes a target at different angles to achieve spatial diversity \cite{Fishler2}. Among these studies, both coherent and non-coherent processing has been considered. Non-coherent processing requires time synchronization between the nodes. Besides time synchronization, coherent processing requires additional phase synchronization  \cite{He1}. Both categories have been shown to offer considerable advantages over conventional radar system in various aspects, such as target detection \cite{He3}, target tracking \cite{Niu,Gorji} and target localization \cite{Dianat,park,Godrich,Liang}. In particular, position information supports an increasing number of location-based applications and services \cite{Win,Shen,Shen2}, therefore target localization is of critical importance for MIMO system.

In general, there are basically two kinds of target localization methods. One is based on the time of arrival (TOA) or angle of arrival (AOA) information from the received signals, which are used to calculate the position via triangulation \cite{Mason,park,Dianat}. Such an algorithm is categorized as an indirect localization approach. The other one, called a direct localization approach, jointly processes the raw signal echos to acquire the maximum-likelihood estimation (MLE) \cite{He1,Niu,Godrich,Greco,Hassanien,BarShalom}. The latter method takes full advantage of received echo information, and thus leads to a higher localization accuracy, especially for weak targets. To obtain the solutions of this method, one of the basic ideas is to employ an iteration algorithm \cite{Rui}, but it requires a proper initial solution from the prior position information, which can restrict the application of this approach in real applications. The other approaches, known as grid-searching methods \cite{Niu}, obtain the target location estimates by searching for the coordinate position that maximizes the likelihood ratio. If only a single target is present, it can be effectively localized using the MLE. However, in many practical situations, there are multiple targets in the coverage area of the system, and multi-target localization is a very challenging problem, for simply expanding the searching dimension to match the number of targets is computationally prohibited.

So far, several problems have been addressed regarding the multi-target localization in radar networks \cite{Gorji,Gogineni,Sun,Kalkan}. In \cite{Gorji}, the multiple-hypothesis (MH)-based algorithm is applied to estimate the number of targets and further achieve the localization for these targets. In \cite{Gogineni} a sparse modeling is proposed for distributed MIMO radar to achieve joint position and velocity estimation of multiple targets. Moreover, motivated by \cite{Gogineni}, \cite{Sun} uses a block sparse Bayesian learning method to estimate the multi-target positions. While in \cite{Kalkan}, the multi-target localization problem is researched using only Doppler frequencies in MIMO radar networks.

Inspired by those works, in this paper, we study the problem of multi-target joint detection and localization for MIMO radar with widely separated antennas. This work is an extension of our previous work \cite{Ai}. Firstly, we present an optimal high dimension localization method based on joint MLE, whose complexity increases exponentially with the number of targets. Besides, without the prior information of the number of targets, a multi-hypothesis testing strategy is required \cite{Schonhoff}, which further complicates this method. To tackle this problem, we then derive two reduced-complexity strategies, specifically, the successive space removal (SSR) algorithm and the successive interference cancellation (SIC) algorithm. The main idea is to split the $2G$ dimensional joint maximization into $G$ disjoint optimization problems. It allows the information of each target to be extracted one by one from the original received signal. It is worth mentioning that our proposed algorithms are based on the threshold decision in detection theory \cite{He4}, hence the target detection information can be simultaneously obtained. In other words, our algorithms belong to a joint multi-target detection and localization procedure, which trades off the algorithm performance for implementation complexity. Numerical examples are provided to assess the detection and localization performances of the our proposed multi-target localization algorithms.

The rest of the paper is organized as follows. The system model is introduced in Section II. In Section III, and the definitions of partially separable and isolated targets are clarified and the high dimensional optimal joint multi-target detection and localization method is derived. In Section IV, two suboptimal algorithms are proposed under the condition that targets are isolated or arbitrarily located, and then the performance of these algorithms is assessed by simulation results in Section V. Finally, Section VI concludes this paper.

\section{Models and Notation}
We assume a typical MIMO radar scenario with $N$ transmitters located at $ (x_k^t,y_k^t),(k = 1,2,...,N)$, and $M$ receivers located at $(x_l^r,y_l^r)$, $(l = 1,2,...,M)$  respectively, in a two-dimensional Cartesian coordinate system. The antennas of both transmitters and receivers are widely separated. A set of mutually orthogonal signals are transmitted, with the lowpass equivalents ${s_k}(t)$, $k = 1,2,...,N $.

The focus in this paper is on simultaneously detecting and localizing multiple targets, therefore only static targets are considered here. Suppose that $G$ (${G \geq 1}$, $G$ is a variable and usually unknown before joint detection and localization) static targets appear in the radar surveillance region, with the ${g}$th  target located at $({x_g},{y_g})$. For convenience, we define a two-dimensional vector ${{\bm{\theta }}_g} \in \mathbb{R}^2$ of the unknown location of the ${g}$th target as
\begin{equation}\label{eq: 1}
\begin{array}{*{20}{c}}
{{\bm{\theta }}_g} \buildrel \Delta \over = {[{x_g},{y_g}]^{\prime}},
\end{array}
\end{equation}
where `` $^\prime$'' denotes the matrix transpose. It should be pointed out that although a 2-dimensional model is adopted here, the extension
to a higher dimensional case is direct.

For noncoherent MIMO radar, the received signal reflected from all ${G}$ targets at the ${l}$th receiver due to the signal transmitted from the ${k}$th transmitter (defined as the ${lk}$th transmit-receive path) is, for $0 < t < T$, given by \footnote{Due to the assumed orthogonality of the signals, it is possible to separate the signal traveling over the ${lk}$th path.}:
\begin{equation}\label{eq: 2}
\begin{split}
{r_{lk}}(t) = & \sum_{g=1}^{G}{ {\alpha _{lkg}}{s_{k}}(t - {\tau _{lkg}})} +{n_{lk}}(t)+{c_{lk}}(t), \\
\end{split}
\end{equation}
where $T$ is the observation time interval. The reflection coefficient ${\alpha _{lkg}} = \left| {{\alpha _{lkg}}} \right|\exp (j{\beta _{lkg}})$ of the ${lk}$th path for the $g$th target is assumed to be a deterministic unknown complex constant with amplitude $\left| {{\alpha _{lkg}}} \right|$ and phase $\beta _{lkg}$ during the observation time $T$. In practice, ${\alpha _{lkg}}$ is related to the Radar Cross Section (RCS) of the $g$th target, and is time varying and unknown before localization in most cases. The term ${\tau _{lkg}}$ denotes the time delay of the received signal from the $g$th target at the $l$th receiver due to the $k$th transmitter, and can be expressed as
\begin{equation}\label{eq: 3}
\begin{split}
&{\tau _{lkg}} =\\
&\frac{{\sqrt {{{({x_g} - x_k^t)}^2} + {{({y_g} - y_k^t)}^2}} {\rm{ + }}\sqrt {{{({x_g} - x_l^r)}^2} + {{({y_g} - y_l^r)}^2}} }}{c},
\end{split}
\end{equation}
with $c$ the speed of light. The terms ${n_{lk}}(t)$ and ${c_{lk}}(t)$ in (\ref{eq: 2}) represents the thermal noise and clutter of the ${lk}$th path. Note that, to accommodate the more general case of moving targets, the signal model with target velocity taken into account can be found in \cite{He1}.

After sampling, the continuous signal of (\ref{eq: 2}) can be written in a vector form
\begin{equation}\label{eq: 4}
{{\bf{r}}_{lk}} = \sum_{g=1}^{G}{ {\alpha _{lkg}}{{\bf{\tilde s}}_{lkg}}} + {{\bf{n}}_{lk}}+ {{\bf{c}}_{lk}},
\end{equation}
where
\begin{equation}\label{eq: 5}
\begin{array}{*{20}{c}}
{{\bf{r}}_{lk}} \buildrel \Delta \over = {[{r_{lk}}[0],{r_{lk}}[1],...,{r_{lk}}[{N_T} - 1]]^{\prime}},
\end{array}
\end{equation}
\begin{equation}\label{eq: 6}
\begin{array}{*{20}{c}}
{{\bf{\tilde s}}_{lkg}} \buildrel \Delta \over = \left[   {{\tilde s_{lkg}}[0],{\tilde s_{lkg}}[1],...,{\tilde s_{lkg}}[{N_T} - 1]}   \right]^{\prime},
\end{array}
\end{equation}
with a sampling interval ${T_s} = {T \mathord{\left/{\vphantom {T {({N_T} - 1)}}} \right.\kern-\nulldelimiterspace} {({N_T} - 1)}}$, thus the sampled signal is $ {r_{lk}}[n] = {r_{lk}}(n{T_s})$, ${\tilde s_{lkg}}[n] = {s_k}(n{T_s} - {\tau _{lkg}})$, $n=0,\ldots,{N_T} - 1$. Note that ${{\bf{\tilde s}}_{lkg}}$ is a function of the unknown target location.
The sampled version of the noise $n_{lk}(t)$ and the clutter $c_{lk}(t)$ in (\ref{eq: 2}), i.e., ${{\bf{n}}_{lk}}$ and ${{\bf{c}}_{lk}}$ in (\ref{eq: 4}), are defined similarly as in (\ref{eq: 5}) as
\begin{equation}
\begin{split}
&{{\bf{n}}_{lk}} \buildrel \Delta \over = {[{n_{lk}}[0],{n_{lk}}[1],...,{n_{lk}}[{N_T} - 1]]^{\prime}},\\
&{{\bf{c}}_{lk}} \buildrel \Delta \over = {[{c_{lk}}[0],{c_{lk}}[1],...,{c_{lk}}[{N_T} - 1]]^{\prime}}.
\end{split}
\end{equation}
The thermal noise and clutter at the $lk$th receive antenna are assumed to be zero-mean complex white Gaussian noise with the correlation matrixes $E\left\{ {{{\bf{n}}_{lk}}{\bf{n}}_{lk}^H} \right\} = \sigma _{lk}^2{\bf{I}}_{N_T}$ and $E\left\{ {{{\bf{c}}_{lk}}{\bf{c}}_{lk}^H} \right\} = {\bf{C}}_{lk}$ respectively, where ${\bf{I}}_d$ denotes the $d \times d$ identity matrix and the superscript `` $\cdot^H$'' denotes conjugate transpose. The temporal correlation matrix of the thermal noise and clutter return is then
\begin{equation}\label{eq: noise_C1}
{\bf{R}}_{lk} = \sigma _{lk}^2{\bf{I}}_{N_T}+ {\bf{C}}_{lk}
\end{equation}
For simplicity, we assume that for a given transmitter-receiver pair, the clutter temporal correlation matrix ${\bf{C}}_{lk}$ is known or estimated a priori. Thus ${\bf{R}}_{lk}$ can be diagonalized by a whitening process. With a slight abuse of notation, we assume such a whitening has been applied prior to (\ref{eq: 4}), but we employ the same notation employed in (\ref{eq: 4}).

Both the thermal noise and clutter echo are assumed to be independent between different transmit-receive paths, thus, for any  $l \ne m$ or $k \ne n$
\begin{equation}\label{eq: noise_C2}
E\left\{ {{{\bf{n}}_{lk}}{\bf{n}}_{mn}^H} \right\} = {{\bf{0}}}, \hspace{5mm} E\left\{ {{{\bf{c}}_{lk}}{\bf{c}}_{mn}^H} \right\} = {{\bf{0}}}. \hspace{5mm}
\end{equation}
This assumption is justified for widely spread antennas.

\section{Joint Multi-target Detection and Localization}
As discussed in \cite{Schonhoff}, the MLE of the unknown parameter vector can be found by examining the likelihood ratio for the hypothesis pair, with ${H_1}$ corresponding to the target presence hypothesis and ${H_0}$ corresponding to the noise only hypothesis. As for multi-target estimation, the observation vector  is related to the parameters of all targets ${{\bm{\theta }}_g},g = 1,2,...,G$. Thus for the joint estimation of all targets, we introduce a high dimensional parameter vector ${\bm{\Theta }}$, which is the concatenation of the individual target parameters, defined as,
\begin{equation}\label{eq: Theta}
{\bm{\Theta }} = {[{\bm{\theta }}_1^{\prime},{\bm{\theta }}_2^{\prime},...,{\bm{\theta }}_G^{\prime}]^{{\prime}}} \in {\mathbb{R}^{2G}}.
\end{equation}

Before proceeding, it is necessary to introduce the following Definition, which is instrumental to the development of the subsequent algorithms.

\emph{\textbf{Definition 1}}: Consider a scenario with $G$ targets and an $M \times N$ MIMO radar. The $g$th and $j$th targets ( $g, j = 1,2,...,G$, and $g \ne j$) are said to be \emph{separable} over the ${lk}$th path, if the time difference of arrival between these two targets is larger than the radar effective pulse width ${\tau _c}$. That means
\begin{equation}\label{eq: 8}
\begin{array}{*{20}{c}}
\left| {{\tau _{lkg}} - {\tau _{lkj}}} \right| > {\tau _{c}},
\end{array}
\end{equation}
where ${\tau_c}$ is the effective duration of the time-correlation of the transmitted waveform ${s_k}(t)$, $k = 1,2,...,N$  \cite{Schonhoff} (for example, if a rectangular pulse with pulse width ${T_p}$ is employed, then ${\tau_c} \simeq {T_p}$). Conversely, the $g$th and $j$th targets are called \emph{inseparable} over the ${lk}$th path if (\ref{eq: 8}) is not satisfied, indicating that the $g$th target shares one range bin in the $lk$th path with the $j$th target. If the $g$th target is \emph{separable} with any other targets in the data plane over all the $M \times N$ transmit-receive paths, the $g$th target is referred to as an \emph{isolated} target. Otherwise, the $g$th target is \emph{partially separable}. Furthermore, if any pairs of targets is mutually \emph{separable} over all paths, then all the $G$ targets are \emph{completely isolated}.

\begin{figure}[h]
\centering
\includegraphics[width=3in]{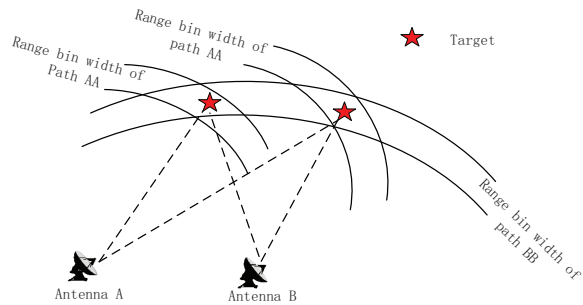}
\caption{Sketch of a scenario with two targets and a $2 \times2$ MIMO radar, wherein the two targets are $inseparable
$ in the $BB$th transmit-receive path.}
\label{fig: 1}
\end{figure}

Take an $M \times N = 2 \times 2$ MIMO radar as an example, where each antenna receives the signals transmitted from other antennas.
A scenario with two \emph{partially separable} targets is plotted in Fig.~\ref{fig: 1} in which only two of the total four paths are plotted. It shows that the two targets are \emph{separable} in the $AA$th propagation path but \emph{inseparable} in the $BB$th path.

\subsection{Optimal High-dimensional Method }
\label{sec:ohm}

In order to simplify the problem, we first assume that the number of targets $G$ is known before localization. Let ${H_1}$ represent the target presence hypothesis as modeled in (\ref{eq: 4}) and ${H_0}$ represents target absence hypothesis, and we can write the likelihood functions of the received vectors of the $lk$th path, i.e., ${{\bf{r}}_{lk}}$, conditioned on the hypotheses and parameters as
\begin{equation}\label{eq: 9}
\begin{split}
 p({{\bf{r}}_{lk}}|{\bm{\Theta }},{{\bm{\alpha }}_{lk}},{H_1}) =& {\kappa_1}\exp \left\{ { - \frac{1}{2}{{\left( {{{\bf{r}}_{lk}} - \sum\limits_{{{g}} = 1}^G { {\alpha _{lkg}}{{{\bf{\tilde s}}}_{lkg}}} } \right)}^H}} \right. \\
&\hspace{0mm} \left. {{\bf{R}}_{lk}^{-1}\left( {{{\bf{r}}_{lk}} - \sum\limits_{{{g}} = 1}^G { {\alpha _{lkg}}{{{\bf{\tilde s}}}_{lkg}}} } \right)} \right\} \\
\end{split}
\end{equation}
and
\begin{equation}\label{eq: 10}
p({{\bf{r}}_{lk}}|{H_0}) = {\kappa_0}\exp \left\{ { - \frac{1}{2}{{\bf{r}}^H_{lk}}{\bf{R}}_{lk}^{ - 1}{{\bf{r}}_{lk}}} \right\},
\end{equation}
where ${{\bm{\alpha }}_{lk}} = {[{\alpha _{lk1}},{\alpha _{lk2}} , \ldots ,{\alpha _{lkG}}]^{\prime}}$ is composed of the unknown complex reflection coefficients of all $G$ targets and ${\kappa_0}$ denotes a constant independent of ${\bm{\Theta }}$. Since $p({{\bf{r}}_{lk}}|{H_0})$ is not a function of ${\bm{\Theta }}$, for the estimation of ${\bm{\Theta }}$, the likelihood function is equivalent to the likelihood ratio \cite{Book1}
\begin{equation}\label{eq: 11}
\begin{split}
\ell({{\bf{r}}_{lk}}{{|}}{\bm{\Theta }},{{\bm{\alpha }}_{lk}}) \propto& \frac{{p({{\bf{r}}_{lk}}{{|}}{\bm{\Theta }},{\bm{\alpha}_{lk}},{H_1})}}{{p({{\bf{r}}_{lk}}|{H_0})}} \\
= &\exp \left\{ {\frac{1}{2}{\bf{r}}_{lk}^H{\bf{R}}_{lk}^{ - 1}\left( {\sum\limits_{{{g}} = 1}^G { {\alpha _{lkg}}{{{\bf{\tilde s}}}_{lkg}}} } \right)} \right. \\
 &\hspace{8mm}+ \frac{1}{2}{\left( {\sum\limits_{{{g}} = 1}^G { {\alpha _{lkg}}{{{\bf{\tilde s}}}_{lkg}}} } \right)^H}{\bf{R}}_{lk}^{ - 1}{{\bf{r}}_{lk}} \\
 &\hspace{8mm}\left. - \frac{1}{{2}}{{\left( {\sum\limits_{{{g}} = 1}^G {{\alpha _{lkg}}{{{\bf{\tilde s}}}_{lkg}}} } \right)}^H}\right.\\
 &\hspace{12mm}\left.{\bf{R}}_{lk}^{ - 1}\left( {\sum\limits_{{{g}} = 1}^G {{\alpha _{lkg}}{{{\bf{\tilde s}}}_{lkg}}} } \right)\right\} .\\
\end{split}
\end{equation}
For any parameter ${\bm{\Theta}}$, the likelihood ratio (\ref{eq: 11}) is maximized using ${{{\bm{\alpha }}_{lk}} = {{{\bm{\hat \alpha }}}_{lk}}}$ \cite{He2}, where ${{{\bm{\hat \alpha }}}_{lk}}$ is calculated as the solution to
\begin{equation}\label{eq: 12}
\frac{\partial }{{\partial {{\bm{\alpha }}_{lk}}}}\ln \ell({{\bf{r}}_{l,k}}{\rm{|}}{\bm{\Theta }},{{\bm{\alpha }}_{lk}}){|_{{{\bm{\alpha }}_{lk}} = {{{\bm{\hat \alpha }}}_{lk }}}} = {\bf{0}}.
\end{equation}
Note that (\ref{eq: 12}) can be written as a group of $G$ equations, with the $g$th ($g=1,2,\ldots, G$) equation expressed as
\begin{equation}\label{eq: lh_1}
\begin{split}
{\bf{\tilde s}}_{lkg}^H{\bf{R}}_{lk}^{ - 1}{\bf{r}}_{lk}^H &-{\alpha _{lkg}}{\bf{\tilde s}}_{lkg}^H{\bf{R}}_{lk}^{ - 1}{{{\bf{\tilde s}}}_{lkg}} \\
&- \sum\limits_{{g_1} = 1,{g_1} \ne g}^G {{\bf{\tilde s}}_{lkg}^H{\bf{R}}_{lk}^{ - 1}} {\alpha _{lk{g_1}}}{{{\bf{\tilde s}}}_{lk{g_1}}} = 0\\
\end{split}
\end{equation}
and the detailed derivation is shown in Appendix A. It can be seen that (\ref{eq: lh_1}) is a linear equation in ${ \alpha _{lk1}},{ \alpha _{lk2}}, \ldots ,{ \alpha _{lkG}}$. Therefore, for compactness, we rewrite the $G$ equations of (\ref{eq: lh_1}) in the following matrix form (also see in Appendix A),
\begin{equation}\label{eq: lh_2}
{\bf{\tilde S}}_{lk}^H{\bf{R}}_{lk}^{ - 1}{{\bf{\tilde S}}_{lk}}{{\bm{\hat \alpha }}_{lk}} = {\bf{\tilde S}}_{lk}^H{\bf{R}}_{lk}^{ - 1}{{\bf{r}}_{lk}},
\end{equation}
with ${{\bf{\tilde S}}_{lk}} = [{{\bf{\tilde s}}_{lk1}},{{\bf{\tilde s}}_{lk2}}, \ldots, {{\bf{\tilde s}}_{lkG}}]$ an ${N_T} \times G$ matrix, and the term ${\bf{\tilde S}}_{lk}^H{\bf{R}}_{lk}^{ - 1}{{\bf{\tilde S}}_{lk}}$ expressed as follows
\begin{equation}\label{eq: lh_1_1}
\left(
  \begin{array}{cccc}
    {\bf{\tilde s}}_{lk1}^H{\bf{R}}_{lk}^{ - 1}{{{\bf{\tilde s}}}_{lk1}} & {\bf{\tilde s}}_{lk1}^H{\bf{R}}_{lk}^{ - 1}{{{\bf{\tilde s}}}_{lk2}} & \cdots & {\bf{\tilde s}}_{lk1}^H{\bf{R}}_{lk}^{ - 1}{{{\bf{\tilde s}}}_{lkG}} \\
    {\bf{\tilde s}}_{lk2}^H{\bf{R}}_{lk}^{ - 1}{{{\bf{\tilde s}}}_{lk1}} & {\bf{\tilde s}}_{lk2}^H{\bf{R}}_{lk}^{ - 1}{{{\bf{\tilde s}}}_{lk2}} & \cdots & {\bf{\tilde s}}_{lk2}^H{\bf{R}}_{lk}^{ - 1}{{{\bf{\tilde s}}}_{lkG}} \\
    \vdots & \vdots & \cdots & \vdots \\
    {\bf{\tilde s}}_{lkG}^H{\bf{R}}_{lk}^{ - 1}{{{\bf{\tilde s}}}_{lk1}} & {\bf{\tilde s}}_{lkG}^H{\bf{R}}_{lk}^{ - 1}{{{\bf{\tilde s}}}_{lk2}} & \cdots & {\bf{\tilde s}}_{lkG}^H{\bf{R}}_{lk}^{ - 1}{{{\bf{\tilde s}}}_{lkG}} \\
  \end{array}
\right).
\end{equation}
If (\ref{eq: lh_1_1}) is invertible (the invertibility of matrix (\ref{eq: lh_1_1}) will be discussed later in this section), using (\ref{eq: lh_2}), we have the ML estimation of ${{\bm{\alpha }}_{lk}}$ as
\begin{equation}\label{eq: lh_3}
%{{\bm{\hat \alpha }}_{lk}} = {{{{\left( {{\bf{\tilde S}}_{lk}^H{\bf{R}}_{lk}^{ - 1}{{{\bf{\tilde S}}}_{lk}}} \right)}^{ - 1}}{\bf{\tilde S}}_{lk}^H{\bf{R}}_{lk}^{ - 1}{{\bf{r}}_{lk}}} \mathord{\left/
% {\vphantom {{{{\left( {{\bf{\tilde S}}_{lk}^H{\bf{R}}_{lk}^{ - 1}{{{\bf{\tilde S}}}_{lk}}} \right)}^{ - 1}}{\bf{\tilde S}}_{lk}^H{\bf{R}}_{lk}^{ - 1}{{\bf{r}}_{lk}}} {\sqrt {\frac{E}{N}} }}} \right.
% \kern-\nulldelimiterspace} {\sqrt {\frac{E}{N}} }}.
{{\bm{\hat \alpha }}_{lk}} = {{{\left( {{\bf{\tilde S}}_{lk}^H{\bf{R}}_{lk}^{ - 1}{{{\bf{\tilde S}}}_{lk}}} \right)}^{ - 1}}{\bf{\tilde S}}_{lk}^H{\bf{R}}_{lk}^{ - 1}{{\bf{r}}_{lk}}}.
\end{equation}

In order to obtain the likelihood of the $lk$th transmit-receive path without parameter ${{\bm{\alpha }}_{lk}}$, we rewrite the logarithmic form of (\ref{eq: 11}) as
\begin{equation}\label{eq: lh_4}
\begin{split}
\ln \ell({{\bf{r}}_{lk}}{{|}}{\bm{\Theta }},{{\bm{\alpha }}_{lk}}) = \frac{1}{2} & \left\{  {  \bf{r}}_{lk}^H{\bf{R}}_{lk}^{ - 1}{{{\bf{\tilde S}}}_{lk}}{{\bm{\alpha }}_{lk}} + {\bm{\alpha }}_{lk}^H{\bf{\tilde S}}_{lk}^H{\bf{R}}_{lk}^{ - 1}{{\bf{r}}_{lk}}\right. \\
&\hspace{2mm}- \left.{\left( {{{{\bf{\tilde S}}}_{lk}}{{\bm{\alpha }}_{lk}}} \right)^H}{\bf{R}}_{lk}^{ - 1}\left( {{{{\bf{\tilde S}}}_{lk}}{{\bm{\alpha }}_{lk}}} \right)\right\}. \\
\end{split}
\end{equation}
Substitution of (\ref{eq: lh_3}) into the third term on the right-hand side of (\ref{eq: lh_4}), we have
\begin{equation}
\begin{split}
&{\left( {{{{\bf{\tilde S}}}_{lk}}{{\bm{\alpha }}_{lk}}} \right)^H}{\bf{R}}_{lk}^{ - 1}\left( {{{{\bf{\tilde S}}}_{lk}}{{\bm{\alpha }}_{lk}}} \right)\\
=&{{\bm{\alpha }}_{lk}}^{H} {{{\bf{\tilde S}}}_{lk}}^{H} {\bf{R}}_{lk}^{ - 1} {{{\bf{\tilde S}}}_{lk}} \left({{\bf{\tilde S}}_{lk}}^{H}{\bf{R}}_{lk}^{ - 1}{{\bf{\tilde S}}_{lk}}\right)^{-1} {{\bf{\tilde S}}_{lk}}^{H} {\bf{R}}_{lk}^{ - 1} {{\bf{r}}_{lk}}\\
=&  {{\bm{\alpha }}_{lk}}^{H} {{{\bf{\tilde S}}}_{lk}}^{H} {\bf{R}}_{lk}^{ - 1} {\bf{r}}_{lk}. \\
\end{split}
\end{equation}
Therefore the summation of the second and third terms on the right-hand side of (\ref{eq: lh_4}) is zero and only the first term remains. Then inserting (\ref{eq: lh_3}) into (\ref{eq: lh_4}), we have
\begin{equation}\label{equ13}
\ln \ell({{\bf{r}}_{lk}}{{|}}{\bm{\Theta }},{{\bm{\alpha }}_{lk}}) =\frac{1}{2}{\bf{r}}_{lk}^H{\bf{R}}_{lk}^{ - 1}{{{\bf{\tilde S}}}_{lk}} ({{\bf{\tilde S}}_{lk}}^{H}{\bf{R}}_{lk}^{ - 1}{{\bf{\tilde S}}_{lk}})^{-1}{{\bf{\tilde S}}_{lk}}^{H}{\bf{R}}_{lk}^{ - 1}{{\bf{r}}_{lk}}. \\
\end{equation}

Due to the independence of observations over different paths, the ML joint detection and estimation of locations of the $G$ targets over all transmit-receive paths can be formulated as
\begin{align}
\label{eq: jml}
&\hspace{-5mm}{{\bm{\hat \Theta }}}_{ML}= \mathop {\arg \max }\limits_{  {\bm{\Theta }} \in {\mathbb{R}^{2G}}} \sum\limits_{k = 1}^N {\sum\limits_{l = 1}^M
\ln \ell({{\bf{r}}_{lk}}{{|}}{\bm{\Theta }},{{\bm{\hat \alpha }}_{lk}})}\\
\label{eq: jml-2}
\text{subject to} \hspace{6mm} & \sum\limits_{k = 1}^N {\sum\limits_{l = 1}^M \ln \ell({{\bf{r}}_{lk}}{{|}}{{{\bm{\hat \Theta }}}_{ML} },{{\bm{\hat \alpha }}_{lk}})}\geq \lambda,
\end{align}
where $\lambda$ is a detection threshold determined by the detection or false alarm probabilities. If the summation of the log-likelihood functions exceeds $\lambda$, a detection of $G$ targets is made, otherwise no target is declared.

Recall that in the beginning of the Section III-A, the number $G$ of targets was assumed to be known before the development of the high-dimensional localization method. The dimension of the multi-target location parameter ${\bm{\Theta }} \in \mathbb{R}^{2G}$ has to be predefined before carrying out the maximization search. If $G$ is unknown, which is the usual case for practical applications, all possible hypotheses of the number of targets have to be evaluated (i.e., a multiple hypotheses testing problem). Owing to the limits of computational complexity, usually an upper bound to the number of prospective targets $G_{\max}$ has to be preset. The number $G_{\max}$ should be set large enough to cover the possibility of the largest number of targets. However a big $G_{\max}$ causes unnecessary computational expense \footnote{Since one has to evaluate all the $G_{\max}$ hypothesis before making a decision,  even if no target is present, $G_{\max}$ searches over the discretized data plane must be performed.} and performance loss due to the increased number of admissible hypotheses.

\subsection{Discussion}
\label{sec:disc}
\subsubsection{The invertibility of matrix (\ref{eq: lh_1_1})} There are cases where (\ref{eq: lh_1_1}) is not invertible. Assume there are $G=2$ targets, then (\ref{eq: lh_1_1}) becomes
\begin{equation}\label{eq: lh_1_12}
\left(
\begin{array}{cc}
  {\bf{\tilde s}}_{lk1}^H{\bf{R}}_{lk}^{ - 1}{{{\bf{\tilde s}}}_{lk1}} & {\bf{\tilde s}}_{lk1}^H{\bf{R}}_{lk}^{ - 1}{{{\bf{\tilde s}}}_{lk2}}\\
  {\bf{\tilde s}}_{lk2}^H{\bf{R}}_{lk}^{ - 1}{{{\bf{\tilde s}}}_{lk1}} & {\bf{\tilde s}}_{lk2}^H{\bf{R}}_{lk}^{ - 1}{{{\bf{\tilde s}}}_{lk2}}\\
  \end{array}
\right).
\end{equation}
If the time delays of the reflected signals from the two targets over the $lk$th path are the same, i.e., $\tau_{lk1}=\tau_{lk2}$, then ${\bf{\tilde s}}_{lk1}^H{\bf{R}}_{lk}^{ - 1}{{{\bf{\tilde s}}}_{lk1}}={\bf{\tilde s}}_{lk2}^H{\bf{R}}_{lk}^{ - 1}{{{\bf{\tilde s}}}_{lk2}}$ and the four elements of (\ref{eq: lh_1_12}) are exactly the same. This means that the rank of the matrix (\ref{eq: lh_1_12}) is one, i.e., (\ref{eq: lh_1_12}) is not invertible, and one can not compute the ML estimation of ${{\bm{\hat \alpha }}_{lk}=}[{\hat \alpha _{lk1}},{\hat \alpha _{lk2}}]$ using (\ref{eq: lh_3}). Actually, when ${\bf{\tilde s}}_{lk1}={\bf{\tilde s}}_{lk2}$, the matrix version of (\ref{eq: lh_2}) is composed of two identical equations from (\ref{eq: lh_1}), thus only one ML estimation of the reflection coefficient can be obtained. This can be explained from a physical point of view, since it is impossible to distinguish and estimate the reflection coefficients for targets with the same time delays over this path. Those cases might be avoided by not looking for targets at these locations, meaning that the search points $(x_1, y_1)$ and $(x_2,y_2)$ which satisfy $\tau_{lk1}=\tau_{lk2}$ are eliminated in the optimization process.

On the contrary, when the time delays of the two targets satisfy $\tau_{lk1} \neq \tau_{lk2}$, the $(1,2)$th and $(2,1)$th elements of (\ref{eq: lh_1_12}) are approximately equal to zero. The $(1,2)$th and $(2,1)$th elements can be viewed as the two reflected signals with different time delays. Thus (\ref{eq: lh_1_1}) is invertible, and two ML estimates of reflection coefficients for each target can be obtained using (\ref{eq: lh_3}). Based on the foregoing discussion, we can see that the invertibility of matrix (\ref{eq: lh_1_1}) relates to the geometric layout of the antennas and targets .

\subsubsection{The curse of dimensionality}
{since no analytic solution exists for the MLE of (\ref{eq: jml}), numerical methods are required. For the grid-search method, in the area of interest ($2G$-dimensional), assume that along the $x$ and $y$ dimensions there are $N_x$ and $N_y$ grid points respectively, implying a total of $(N_x \times N_y)^{G}$ grid points. The unit size of each dimension is chosen based on the characteristics of radar system (e.g., range resolution), the geographical setting of the radar antennas with respect to the area of interest and the computational resources. After the grid search, standard optimization methods can also be employed to refine the estimation \cite{Niu}. Although the grid-search implementation of (\ref{eq: jml})  is straightforward in principle, it involves a high-dimensional joint maximization. Since the discretized data plane contains $N_x \times N_y$ grid cells, the total complexity increases exponentially with the number of targets $G$. Therefore this high-dimensional multi-target localization method is computationally prohibitive if there are more than a few targets. }

%\subsubsection{Multi-hypothesis testing problem}
%recall that in the beginning of the Section III-A, the number $G$ of targets is assume to be known before the development of the high-dimensional localization method. The dimension of the multi-target location parameter ${\bm{\Theta }} \in \mathbb{R}^{2G}$ has to be predefined before carrying out the maximization search. If target number $G$ is unknown, which is the usual case for practical applications, all possible hypotheses of the number of target have to be evaluated (i.e., a multiple hypotheses testing problem \cite{Schonhoff}). Owing to the limits of computational complexity, usually an upper bound to the number of prospective targets $G_{\max}$ has to be preset. The number $G_{\max}$ should be set large enough to cover the possibility of the largest number of targets. However a big $G_{\max}$ causes unnecessary computational expense \footnote{Since one has to evaluate all the $G_{\max}$ hypothesis before making a decision,  even if no target is present, $G_{\max}$ number of searches over the discretized data plane has to be performed.} and performance loss due to the increased number %of admissible hypotheses.

The above problems and the multi-hypothesis testing problem mentioned before heavily restrict the applications of the high-dimensional method. Hence, suboptimal algorithms are also investigated in the subsequent sections to trade off algorithm performance for implementation complexity.

\section{Suboptimum Strategies}
\subsection{Successive-Space-Removal Algorithm}
The aim of this subsection is to derive reduced-complexity strategies for implementing the MLE (\ref{eq: jml}), at the price of estimation performance tradeoff. The main idea is to split the $2G$-dimensional joint maximization into $G$ disjoint optimization problems, which allows information about each target to be extracted one-by-one from the original received signal.

The design of this suboptimal algorithm is based upon the assumption that the targets present in the radar surveillance region are completely \emph{isolated}. Normally a MIMO radar receiver incorporates thousands of resolution range bins, so completely isolated targets are not rare. In this case, from \emph{Definition 1} and the fact that ${\bf{R}}_{lk}^{ - 1}$ is a diagonal matrix, for any $g,j=1,\ldots,G$ and $g\neq j$, as discussed in Section \ref{sec:disc}, ${{{\bf{\tilde s}}}_{lkg}}$ and ${{{\bf{\tilde s}}}_{lkj}}$ corresponding to the $g$th and $j$th target respectively must effectively meet the condition
\begin{equation}\label{eq: premise}
{\bf{\tilde s}}_{lkg}^H {\bf{R}}_{lk}^{- 1} {\bf{\tilde s}}_{lkj} = 0.
\end{equation}
Thus the matrix ${\bf{\tilde S}}_{lk}^H{\bf{R}}_{lk}^{ - 1}{{\bf{\tilde S}}_{lk}}$ is diagonal, and then the closed-form ML estimation of ${\alpha _{lkg}}$ is obtained, by using (\ref{eq: lh_1}), as
\begin{equation}\label{eq: alpha}
{\hat \alpha _{lkg}} = \frac{   {{{{\bf{\tilde s}}}^H}_{lkg}{\bf{R}}_{lk}^{ - 1}{{\bf{r}}_{lk}}}   }  {  {\bf{\tilde s}}_{lkg}^H{\bf{R}}_{lk}^{ - 1}{{{\bf{\tilde s}}}^{}}_{lkg}}.
\end{equation}
By substituting (\ref{eq: alpha}) back into (\ref{eq: lh_4}), we have
\begin{equation}\label{equ15}
\begin{split}
\ln \ell({{\bf{r}}_{lk}}{{|}}{\bm{\Theta }},{{\bm{\hat \alpha }}_{lk}})&=\frac{1}{2} {\bf{r}}_{lk}^H{\bf{R}}_{lk}^{ - 1}\sum\limits_{g= 1}^G{\hat\alpha _{lk{g}}}{\bf{\tilde s}}_{lkg}\\
&=\frac{1}{2} \sum\limits_{g= 1}^G {\bf{r}}_{lk}^H {\bf{R}}_{lk}^{ - 1} {\bf{\tilde s}}_{lkg} \frac{ {{{{\bf{\tilde s}}}^H}_{lkg}{\bf{R}}_{lk}^{ - 1}{{\bf{r}}_{lk}}}   }  {{ {\bf{\tilde s}}_{lkg}^H{\bf{R}}_{lk}^{ - 1}{{{\bf{\tilde s}}}^{}}_{lkg}}}\\
&=\frac{1}{2} \sum\limits_{g= 1}^G \frac{ 1 }  {{ {\bf{\tilde s}}_{lkg}^H{\bf{R}}_{lk}^{ - 1}{{{\bf{\tilde s}}}^{}}_{lkg}}} {|{{{{\bf{\tilde s}}}^H}_{lkg}{\bf{R}}_{lk}^{ - 1}{{\bf{r}}_{lk}}}|^{ 2 }}. \\
\end{split}
\end{equation}
Substitution of (\ref{equ15}) into (\ref{eq: jml}) gives
\begin{equation}\label{eq: jml_0}
\begin{split}
&{{\bm{\hat \Theta }}_{ML}}= \mathop {\arg \max }\limits_{\left({{\bm{\theta }}_1}, \cdots, {{\bm{\theta }}_{{G}}}\right) \in {\mathbb{R}^{2G}}} \sum\limits_{k = 1}^N \sum\limits_{l = 1}^M \sum\limits_{g = 1}^G { \ell_{lk}({{\bm{\theta }}_g})}\\
&\hspace{8mm} =\mathop {\arg \max }\limits_{\left({{\bm{\theta }}_1}, \cdots, {{\bm{\theta }}_{{G}}}\right) \in {\mathbb{R}^{2G}}} \sum\limits_{g = 1}^G { \mathcal{F}({{\bm{\theta }}_g})},\\
\text{subject to}&\hspace{9mm}\text {${{\bm{\hat \theta }}_1}, \cdots, {{\bm{\hat \theta }}_{{G}}}$ are comletely \emph{islated}.}\\
\end{split}
\end{equation}
where
\begin{equation}\label{eq: ll_lk}
{ \ell_{lk}({\bm{\theta }_g})}=\frac{1}{2} {\frac{1}{{{\bf{\tilde s}}_{lkg}^H{\bf{R}}_{lk}^{ - 1}{{{\bf{\tilde s}}}_{lkg}}}}{{\left| {{\bf{\tilde s}}_{lkg}^H{\bf{R}}_{lk}^{ - 1}{{\bf{r}}_{lk}}} \right|}^2}}
\end{equation}
is the log-likelihood function for a single target location ${\bm{\theta }_g}$ for the $lk$th transmit-receive path, and
\begin{equation}\label{eq: jml_f}
\mathcal{F}({{\bm{\theta }_g}}) = \sum\limits_{k = 1}^N {\sum\limits_{l = 1}^M {\ell_{lk}({{\bm{\theta }_g}})} }
\end{equation}
is defined as the objective function of the $g$th single target location ${\bm{\theta }_g}$. The right-hand side of (\ref{eq: ll_lk}) implies that $\ell_{lk}({\bm{\theta }_g})$ will be large only when ${\bf{r}}_{lk}$ can be well matched with ${\bf{\tilde s}}_{lkg}$.

For the scenario with completely isolated targets, the maximum of the summation of $G$ objective functions in (\ref{eq: jml_0}) is equal to the summation of $G$ maximums of the objective functions, because this special scenario excludes the case where two targets are are in a common range bin for any path. According to this fact, we can reasonably simplify the high-dimensional optimization problem in (\ref{eq: jml_0}) by reducing the dimension of the search space. So (\ref{eq: jml_0}) can be approximately expressed as
\begin{equation}\label{eq: jml_01}
{{\bm{\hat \Theta }}_{ML}} = [\bm{\hat \theta}_{1}^{\prime},\bm{\hat \theta}_{2}^{\prime},\cdots ,\bm{\hat \theta}_{G}^{\prime}]^{\prime}
\end{equation}
with its $g$th element estimated as
\begin{equation}\label{eq: 30}
{{{\bm{\hat \theta }}_g}=\mathop {\arg \max }\limits_{ {{\bm{\theta }}_g} \in {\mathbb{S}_{g}  } } \mathcal{F}({{\bm{\theta }}_g})},
\end{equation}
where the initial parameter space ${\mathbb{S}_{1}}=\mathbb{R}^2$, and for $g=2,\ldots,G$,
\begin{equation}\label{eq: S}
{\mathbb{S}_{g}}=\mathbb{S}_{g-1}\setminus {{{{\mathbb{B}}}}(\bm{\theta}, {{\bm{\hat \theta }}_{g-1}})}. \footnote{The symbol $A \setminus B$ denotes the set difference of set $A$ and $B$.}
\end{equation}
${{{{\mathbb{B}}}}(\bm{\theta}, {{\bm{ \theta }}_g})}$, written succinctly as ${{{{\mathbb{B}}}}({{\bm{ \theta }}_g})}$, is defined as that subset of the search area, which includes the range bins for those paths which are occupied (see Fig.~\ref{fig: 1}) by the target located at ${{\bm{ \theta }}_g}$, which is written as
\begin{equation}\label{eq: S1}
{{\mathbb{B}}({{\bm{ \theta }}_g})}=\bigcup_{l=1}^M\bigcup_{k=1}^N {{{{\mathbb{B}}}}_{lk}(\bm{\theta}, {{\bm{ \theta }}_g})},
\end{equation}
where ${{{{\mathbb{B}}}}_{lk}(\bm{\theta},{{\bm{ \theta }}_g})}$, similarly succinctly written as ${{{{\mathbb{B}}}}_{lk}({{\bm{ \theta }}_g})}$, denotes the part of ${{\mathbb{B}}({{\bm{ \theta }}_g})}$ corresponding to the $lk$th path.

In order to solve the optimization problem, we need to accurately find the maximums of each objective function $\mathcal{F}({{\bm{\theta }_g}})$ constrained by different conditions. The estimator (\ref{eq: 30}) can provide a computational efficient and practical method to find the maximums at the scenario with completely isolated targets. From the previous analysis, we can know two targets are not present in the area defined in (\ref{eq: S}), indicating no target shares a common range bin. In fact, this is actually implied by the constraint corresponding to the optimization problem in (\ref{eq: jml_01}) and (\ref{eq: 30}). It also means that it is reasonable to find each maximum one by one by eliminating the areas corresponding to every determined target.

Because the true location ${\bm{\theta}}_g$ corresponding to the $g$th target is unknown, we need to substitute the estimation result ${{\bm{\hat \theta }}_g}$ for ${{\bm{ \theta }}_g}$. Taking the potential estimation error between ${{\bm{\hat \theta }}_g}$ and the true position ${{\bm{ \theta }}_g}$  into consideration, the correctness of the decision of range bins in each path is not guaranteed. Thus, in implementation, ${{{{\mathbb{B}}}}_{lk}({{\bm{\hat \theta }_g}})}$ is defined as follows (set the error margin as a range bin)
\begin{equation}\label{eq: blk}
\begin{split}
%{{{{\mathbb{B}}}}_{lk}({{\bm{\hat \theta  }_g}})} = \left\{ {\left( {x,y} \right)| {\left\lfloor {{{\tau _{lkg}} \mathord{\left/
% {\vphantom {{{\tau _{lk}}\left( {x,y} \right)} {{\tau _c}}}} \right.
% \kern-\nulldelimiterspace} {{\tau _c}}}} \right\rfloor  - \left\lfloor {{{{{\hat \tau }_{lk{g}}}} \mathord{\left/
% {\vphantom {{{{\hat \tau }_{lk{g}}}} {{\tau _c}}}} \right.
% \kern-\nulldelimiterspace} {{\tau _c}}}} \right\rfloor }  \le 1} \right\},
{{{{\mathbb{B}}}}_{lk}({{\bm{\hat \theta  }_g}})} = \{\left( {x,y} \right)| \lfloor {\tau _{lkg}}(x,y)/{{\tau _c}}\rfloor - \lfloor \hat{\tau}_{lkg}/{{\tau _c}}\rfloor \leq 1 \},
 \end{split}
\end{equation}
where
\begin{equation}\label{eq: 29_1}
\begin{split}
{\hat \tau _{lk{g}}} =& \frac{1}{c}\left( {\sqrt {{{(\hat{x} - x_k^t)}^2} + {{(\hat{y} - y_k^t)}^2}}} \right.\\
 & \hspace{20mm}+ \left.{\sqrt {{{(\hat{x} - x_l^r)}^2} + {{(\hat{y} - y_l^r)}^2}} } \right)\\
 \end{split}
\end{equation}
is the time delay of the estimated target located at ${{\bm{\hat \theta }}_{{g}}}$  in the $lk$th path, ${\tau _c}$ is the effective duration of the time-correlation function of the transmitted waveform and $\left\lfloor a \right\rfloor$ is the maximum integer not greater than $a$. Therefore, ${\mathbb{S}_{g}}$ in (\ref{eq: S}) represents the search space after removing the area affected by the first $g-1$ declared targets.

A variant of (22) with much lower complexity can be expressed as follows
\begin{equation}\label{eq: jml_v}
\begin{split}
&\hspace{3mm}{{\bm{\hat \Theta }}_{ML}} = [\bm{\hat \theta}_{1}^{\prime},\bm{\hat \theta}_{2}^{\prime},\cdots ,\bm{\hat \theta}_{G}^{\prime}]^{\prime}\\
& \hspace{8mm}\\
\text{with}& \hspace{8mm} {{\bm{\hat \theta }}_g}  = \mathop {\arg \max }\limits_{ {{\bm{\theta }}_g} \in {\mathbb{S}_{g}}} \mathcal{F}({{\bm{\theta }}_g})\\
\text{subject to}& \hspace{8mm} \sum\limits_{g = 1}^G { \mathcal{F}({{\bm{\theta }}_g})} \geq \lambda.\\
\end{split}
\end{equation}
As is mentioned before, the structure of (\ref{eq: jml_v}) indicates that the 2$G$-dimensional maximization of (\ref{eq: jml_0}) can be replaced by sequentially implementing $G$ 2-dimensional maximizations, i.e., finding the ${{\bm{\hat \theta }}_g} \in {\mathbb{S}_{g}}$, $g=1,\ldots, G$, which maximize $\mathcal{F}({{\bm{\theta }}_g})$, then removing the search area affected by ${{\bm{\hat \theta }}_g} $ to form the search space ${\mathbb{S}_{g+1}}$ for the next maximization until $g=G$. By doing this, the complexity is reduced significantly, and we refer to this algorithm as the successive-space-removal (SSR) multi-target localization method.

However, SSR would also face the cumbersome multi-hypothesis testing problem when target number $G$ is unknown. To deal with this, we propose a step-by-step detection procedure for SSR. Since the existence of a certain target is irrelevant to other targets under the assumption that the targets are \emph{completely isolated}, we can approximately replace the detection process in (\ref{eq: jml_v}) with $G$ single target detection problems as
\begin{equation}\label{eq_33}
\mathcal{F}({{\bm{\hat \theta }}_g}) {\mathop{\gtrless}\limits^{H_1}_{H_0}} \lambda_g,\   \ \ \ g = 1, 2, \ldots, G,
\end{equation}
where ${\lambda _g}$ is the threshold of the $g$th single target detection process. Usually threshold ${\lambda _g}$ is chosen to achieve a certain false alarm probability. If the background is homogeneous, one can use the same threshold $\lambda'$ for all $G$ detection processes. In cases where the number $G$ of targets is not available, the localization process can be terminated automatically if the $G^\prime$th estimated location ${{\bm{\hat \theta }}_{G^\prime}}$ is determined as not target, i.e., $\mathcal{F}({{\bm{\hat \theta }}_{G^\prime}}) < \lambda'$. This simply relies on the fact that $\mathcal{F}({{\bm{\theta }}_{{G^\prime} + 1}}) \le \mathcal{F}({{\bm{\hat \theta }}_{{G^\prime}}}) < \lambda '$ when the background is homogeneous, meaning that every estimate in the subsequent search will be decided as $H_0$. Since the threshold $\lambda'$ remains the same on the whole data plane in each detection process, the decision of the threshold for all $G$ detection processes is made only once to narrow down the possible locations of the targets. A summary of the proposed SSR algorithm under homogeneous background is given in Algorithm \ref{algorithm: SSR}. It should be noted that for the non-homogeneous environment, in order to achieve the desired constant false alarm rate, the value of detection threshold in (\ref{eq: CC1}) needs to be adapted along with the variety of the noise/clutter, i.e., false alarm rate approach \cite{Fuhrmann, Rohling, He4}. Besides, in Algorithm \ref{algorithm: SSR}, we set an upper bound $G_{\max}$ for the maximum number of the potential targets. Thus when $G_{\max}$ estimated locations have been obtained, the iteration ends automatically to avoid the overload of the system.
\LinesNumbered
%\linesnumbered
\begin{algorithm}[htb]\label{algorithm: SSR}
%\dontprintsemicolon
\DontPrintSemicolon
%\SetAlgoLined
\caption{\label{alg:summary} {The Summary of SSR Algorithm}}
Compute the objective function $\mathcal{F}({\bm{\theta }})$ for the parameter space of interest ${{\bm{\theta }}} \in \mathbb{R}^2$.\;
Form the original set ${\Phi}_1$ of the estimated candidates as
\begin{equation}\label{eq: CC1}
 {\Phi_1}= \left\{ {\bm{\theta }}: \mathcal{F}({{\bm{\theta }}}) > {\lambda ^{\prime}}, {{\bm{\theta }}} \in \mathbb{R}^2\right\}.
 \end{equation}
and the set of localized targets ${\Omega _D} = \emptyset $.

\For{ $g=1,2,\ldots, G_{\max}$ }{
Obtain the $g$th maximum likelihood estimate as
$
{{\bm{\hat \theta }}_g} = \arg \max_{{{\bm{\theta }}} \in {\Phi_g}} \mathcal{F}({\bm{\theta }}).
$\;
Add the $g$th estimate ${{\bm{\hat \theta }}_g}$ to the set ${\Omega}_D$ of the declared targets, i.e., ${\Omega}_D=\{{\bm{\hat \theta }}_1,\ldots,{\bm{\hat \theta }}_g \}$.\;
Update the estimate candidate set ${\Phi}_g$ by subtracting the set ${\Psi}({{\bm{\hat \theta }}_g})$, whose elements share common range bins with ${{\bm{\hat \theta }}_g}$, as
\begin{equation}\label{eq: 351}
{\Phi_{g+1}}= {\Phi}_{g} \ /{\Psi}({{\bm{\hat \theta }}_g}),
\end{equation}
where ${\Psi}({{\bm{\hat \theta }}_g})=\{ {\bm{\theta }}: {\Phi}_g \cap \mathbb{B}({{\bm{\hat \theta }}_g}) \}$.  \;
\If{${\Phi_{g+1}} =  \emptyset$ or $g+1>G_{\max}$ }
{end the $\mathbf{for}$ loop.\;}
}
Output the set ${\Omega}_D$ containing the locations of the detected target, and the number of elements of the set ${\Omega}_D$ is the number of targets. \;
\end{algorithm}

%$Sil(K) = \sum\nolimits_{l = 1}^K {\sum\nolimits_{p = 1}^{{n_l}} {\left( {{{\left( {{b_{lp}} - {a_{lp}}} \right)} \mathord{\left/
% {\vphantom {{\left( {{b_{lp}} - {a_{lp}}} \right)} {\max \left\{ {{a_{lp}},{b_{lp}}} \right\}}}} \right.
% \kern-\nulldelimiterspace} {\max \left\{ {{a_{lp}},{b_{lp}}} \right\}}}} \right)} } $

\begin{figure}[!t]
\centering
\subfigure[]{\includegraphics[width=6.5cm]{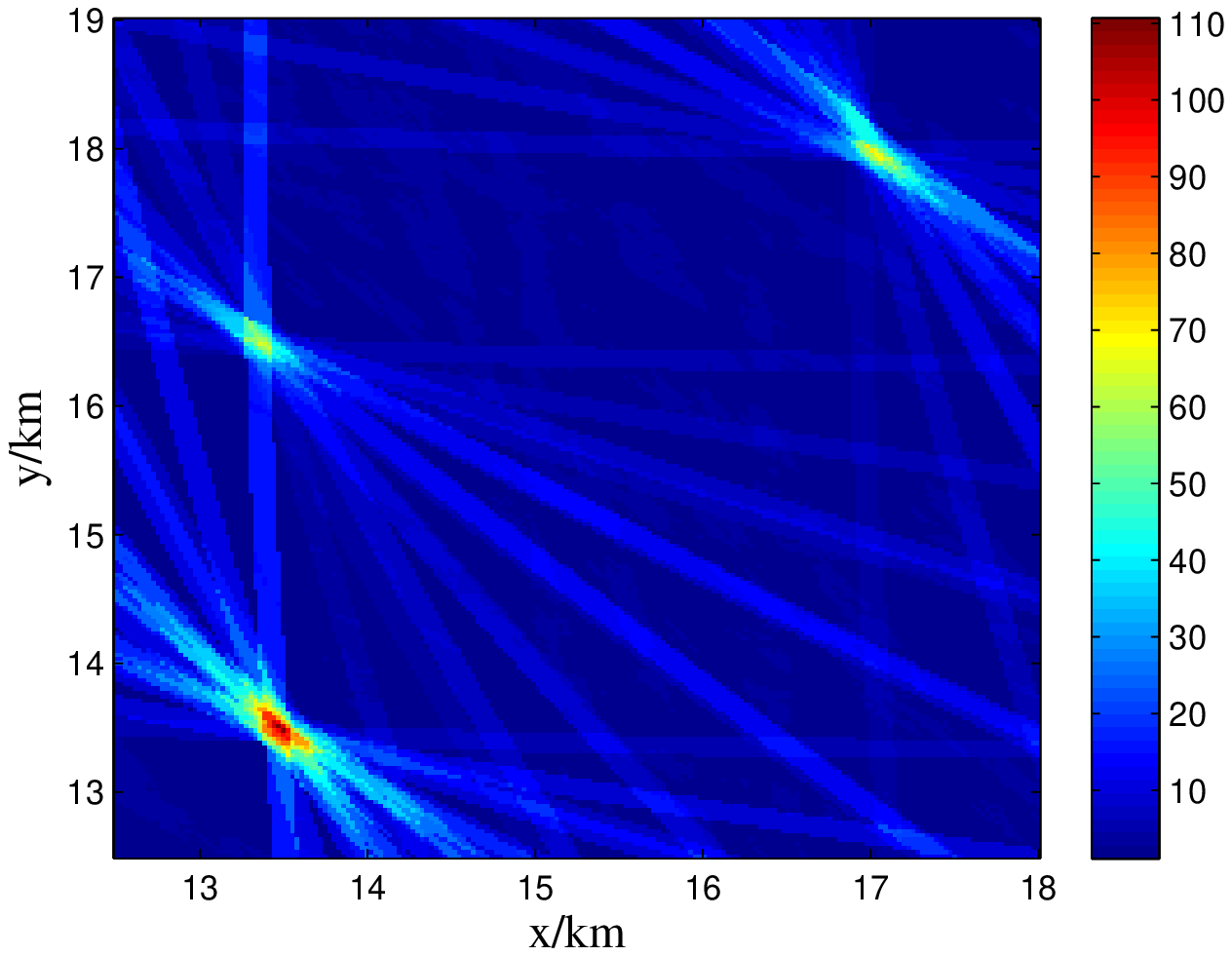}}
\subfigure[]{\includegraphics[width=6.5cm]{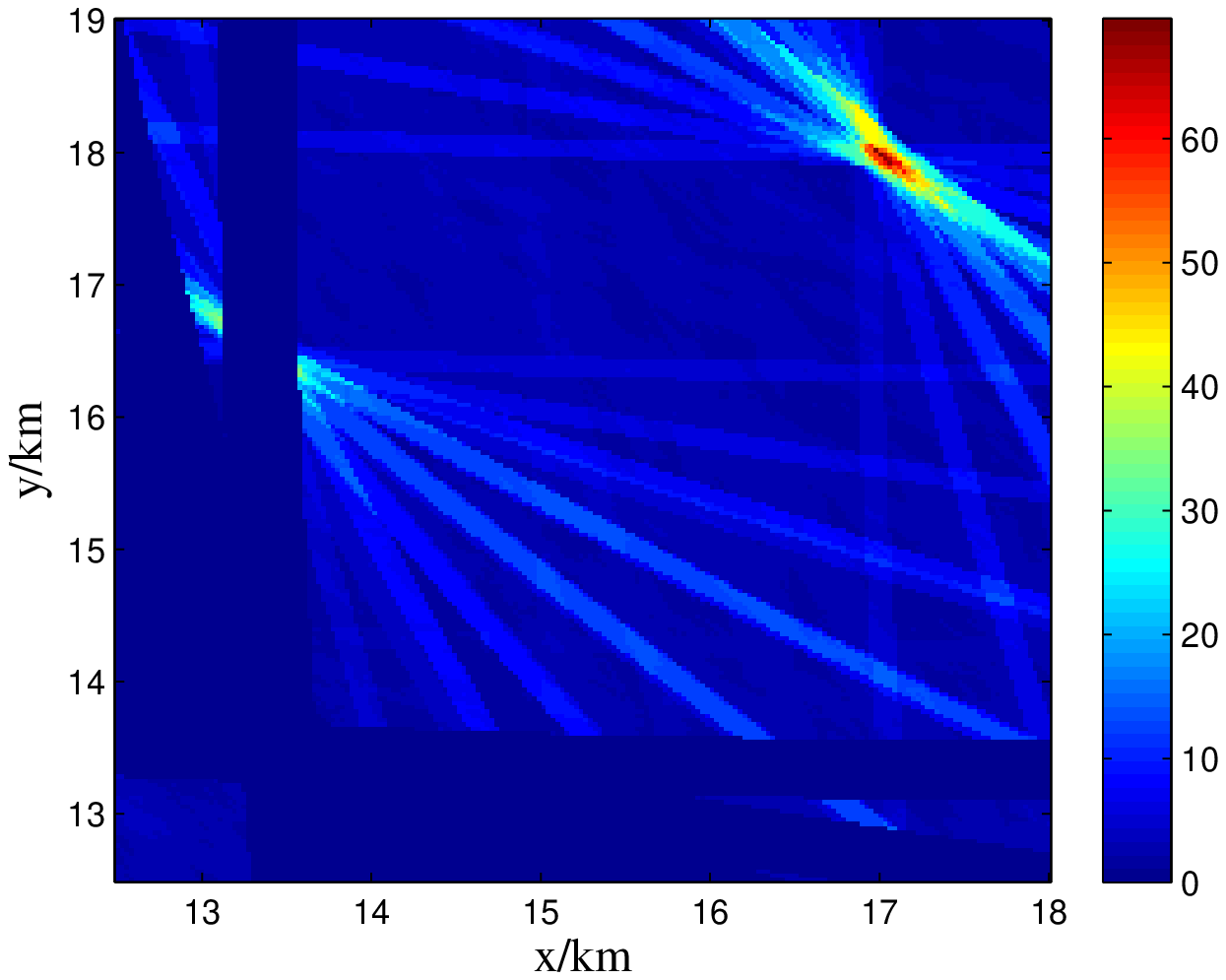}}
\caption{Illustrations of the objective function for a scenario with three targets wherein the two targets on the left-hand side are \emph{inseparable} (a) the original objective function $\mathcal{F}({\bm{\theta }})$. (b) the objective function after the removal of the search space related to the detected target on the lower left-hand corner.}
\label{fig: ssr}
\end{figure}

When the assumption that all targets are \emph{completely isolated} holds, SSR can sequentially localize multiple targets efficiently with no need for a multi-hypotheses testing algorithm. However, for more general cases, targets located arbitrarily may share range bins with each other in one or more transmit-receive paths, i.e., \emph{partially separable}. In this case, the direct removal of the search space of detected targets using (\ref{eq: S}) and (\ref{eq: S1}) will result in the miss-detection of subsequent targets which are \emph{inseparable} over certain pathes with the previously detected targets. Fig. \ref{fig: ssr}(a) shows a scenario with three targets wherein the two targets on the left-hand side are \emph{inseparable}. It can be seen in Fig. \ref{fig: ssr}(b) that the elimination of the area corresponding to the target on the lower left-hand corner (stronger one) will hinder the subsequent detection and localization of the target on the upper left-hand side. In order to deal with this problem, a carefully designed suboptimal strategy is given in the next subsection.

 %is suboptimal in the sense that

%\begin{figure}[h]
%\begin{minipage}[b]{0.48\linewidth}
%  \centering
%  \centerline{\includegraphics[width=4.8cm]{SSRfig1.eps}}
%%  \vspace{1.5cm}
%  \centerline{(a)}\medskip
%\end{minipage}
%\hfill
%\begin{minipage}[b]{0.48\linewidth}
%  \centering
%  \centerline{\includegraphics[width=4.8cm]{SSRfig2.eps}}
%%  \vspace{1.5cm}
%  \centerline{(b) }\medskip
%\end{minipage}
%\caption{Mesh plot of the objective function for a scenario with three targets wherein the two targets on the right-hand side are \emph{unseparable} (a) the original objective function $\mathcal{F}({\bm{\theta }})$. (b) the objective function after the removal of the search space related to the detected target on the lower left-hand corner.}
%\label{fig: ssr}
%\end{figure}

%The assumption that targets present are always \emph{completely separable} is too stringent for realistic applications.

\subsection{Successive-Interference-Cancellation Algorithm}

The new algorithm differs from SSR in that it does not directly clear search space affected by the targets detected as in (\ref{eq: S}) and instead only eliminates the interference of the extracted targets from the objective function. As a consequence, the objective function changes every time after a target is detected. In this way, another variant of (\ref{eq: jml}) for the ML joint detection and localization of multiple targets can be formulated as
\begin{equation}\label{eq: jml_v2}
\begin{split}
&\hspace{3mm} {{\bm{\hat \Theta }}_{ML}} = [\bm{\hat \theta}_{1}^{\prime},\bm{\hat \theta}_{2}^{\prime},\cdots ,\bm{\hat \theta}_{G}^{\prime}]^{\prime}\\
&\hspace{3mm}\\
\text{with}& \hspace{8mm}  {{{\bm{\hat \theta }}_g}=\mathop {\arg \max }\limits_{ {{\bm{\theta }}_g} \in {\mathbb{R}^{2}  } } \mathcal{F}_{g}({{\bm{\theta }}_g})}\\
\text{subject to}& \hspace{8mm} \sum\limits_{g = 1}^G { \mathcal{F}_g({{\bm{\theta }}_g})}\geq \lambda.\\
\end{split}
\end{equation}
where $\mathcal{F}_g({{\bm{\theta }}})$ is the objective function for the $g$th maximization (i.e., extraction of the $g$th target) and is defined as follows
\begin{equation}\label{eq: m_1}
\begin{split}
\mathcal{F}_{g+1}({{\bm{\theta }}})&=\mathcal{F}_g({{\bm{\theta }}})-\sum_{k=1}^{N}\sum_{l=1}^{M}\mathcal{M}_{lkg}({{\bm{\theta }}})\\
&=\mathcal{F}({{\bm{\theta }}})-\sum_{i=1}^{g}\sum_{k=1}^{N}\sum_{l=1}^{M}\mathcal{M}_{lki}({{\bm{\theta }}}),\\
\end{split}
\end{equation}
with the term ${\mathcal{M}_{lkg}}({{\bm{\theta }}})$ in (\ref{eq: m_1}) referred to as the modified term related to the ${g}$th detected target over the $lk$th path. In order to eliminate the interference to the likelihood from the previously detected targets, the modified term of the ${g}$th detected target over the $lk$th path is defined as
\begin{comment}
\begin{equation}\label{eq: 28}
{\mathcal{M}_{lkg}}({{\bm{\theta }}}) =
\left\{
\begin{array}{cc}
\ell_{lk}({{\bm{\hat \theta }}}_g),   &  {{\bm{\theta }}} \in \mathbb{B}_{lk}({{\bm{\hat \theta }}}_g)\setminus \mathbb{B}_{lk}({{\bm{\hat \theta }}}_{g-1})\cdots \setminus \mathbb{B}_{lk}({{\bm{\hat \theta }}}_1) \\
 0,  &  {{\bm{\theta }}} \notin \mathbb{B}_{lk}({{\bm{\hat \theta }}}_g)\setminus \mathbb{B}_{lk}({{\bm{\hat \theta }}}_{g-1})\cdots \setminus \mathbb{B}_{lk}({{\bm{\hat \theta }}}_1), \\
 \end{array}
 \right.
\end{equation}
\end{comment}
\begin{equation}
{\mathcal{M}_{lkg}}({{\bm{\theta }}}) =
\left\{
\begin{split}
& \ell_{lk}({{\bm{ \theta }}}),\hspace{4.5mm}{{\bm{\theta }}} \in \mathbb{B}_{lk}({{\bm{\hat \theta }}}_g)\setminus \mathbb{C}_{lk}({{\bm{\hat \theta }}}_{g})\\
& 0, \quad\quad \:\,\hspace{3mm} \text{otherwise}
\end{split}
\right.
\end{equation}
with
\begin{equation}
\mathbb{C}_{lk}({{\bm{\hat \theta }}}_{g})=\mathbb{B}_{lk}({{\bm{\hat \theta }}}_g) \cap \left\{\mathbb{B}_{lk}({{\bm{\hat \theta }}}_1)\cup \cdots\cup \mathbb{B}_{lk}({{\bm{\hat \theta }}}_{g-1})\right\}
\end{equation}
where the terms $\ell_{lk}({{\bm{ \theta }}}) $ and  $\mathbb{B}_{lk}({{\bm{ \theta }}})$ are the defined by (\ref{eq: ll_lk}) and (\ref{eq: blk}) respectively.
In essence, the modified term ${\mathcal{M}_{lkg}}$ is equal to the log-likelihood over the $lk$th transmit-receive path for the parameter space that is affected by the estimated target ${{\bm{\hat \theta }}}_g$, i.e., $\mathbb{B}_{lk}({{\bm{\hat \theta }}}_g)$, otherwise it is zero. However, for a certain parameter ${{\bm{\tilde \theta }}} \in \mathbb{B}_{lk}({{\bm{\hat \theta }}}_g)$, its log-likelihood over the $lk$th path may have already been subtracted in the previous modifications of the objective function, namely, ${{\bm{\tilde{\theta} }}} \in \left\{\mathbb{B}_{lk}({{\bm{\hat \theta }}}_1)\cup \cdots\cup \mathbb{B}_{lk}({{\bm{\hat \theta }}}_{g-1})\right\}$. Hence, ${\mathcal{M}_{lkg}}$ is equal to $\ell_{lk}({{\bm{ \theta }}})$ only for the parameter space ${{\bm{\theta }}} \in \mathbb{B}_{lk}({{\bm{\hat \theta }}}_g)\setminus \mathbb{C}_{lk}({{\bm{\hat \theta }}}_{g})$, otherwise zero.

Then ${\mathcal{F}_g}({{\bm{\theta }}}) $ can be viewed as a modified form of the original objective function $\mathcal{F}({{\bm{\theta }}})$, wherein the likelihood interference from the previously detected $g-1$ targets has been eliminated. Hence we refer to this algorithm as a successive-interference-cancellation (SIC) algorithm. The idea of SIC is similar in spirit to the well known CLEAN algorithm \cite{Clark}. To further reduce complexity, it should be noted that the log-likelihood values of all the $MN$ paths have already been calculated when we compute the original objective function $\mathcal{F}({{\bm{\theta }}})$. So there is no need to recalculate the log-likelihood values to generate the modified term.

\begin{figure}[!t]
\centering
\centerline{\includegraphics[width=6.8cm]{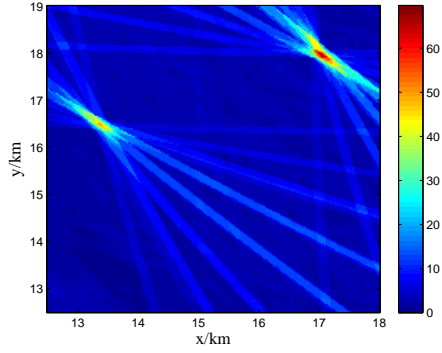}}
\caption{An illustration of the modified objective function for the same scenario as in Fig. \ref{fig: ssr}(a). The interference related to the detected target on the lower left-hand corner has been subtracted from the original objective function.}
\label{fig: sic}
\end{figure}

It can be seen that (\ref{eq: jml_v}) and (\ref{eq: jml_v2}) have exactly the same structure. Therefore, similar to the implementation of SSR, SIC can also be performed sequentially to break down the high-dimensional joint maximization and  avoid the multiple hypotheses testing problem. Nevertheless, there are two differences between SSR and SIC. Firstly, for each iteration, SIC only modifies the objective function to clear the interference of detected targets and keeps the search space intact, rather than deleting the search area as in SSR. This greatly facilitates the detection and localization of \emph{inseparable} targets. We still consider the same scenario shown in Fig. \ref{fig: ssr}(a) wherein the two targets in the left-hand side are \emph{inseparable}. The modified objective function after eliminating the interference of the target on the lower left-hand corner (stronger one) using SIC is shown in Fig. \ref{fig: sic}.
It can be seen that compared to the objective function in Fig. \ref{fig: ssr}(b), SIC is able to reserves more information of the target on the upper left-hand side (\emph{inseparable} with the detected and located one), making the subsequent detection and localization of this target possible.

Secondly, the setting of the detection threshold for SIC is more complicated. The reason is that for different parts of the parameter space, the modified objective function ${\mathcal{F}_g}({{\bm{\theta }}})$ defined in (\ref{eq: m_1}) is composed of the likelihood summation of different number of paths. Thus even for the homogeneous background, the value of the detection threshold may change for different parts of the parameter space to prevent missing targets. For this reason, we define the detection threshold of the parameter ${{\bm{\theta }}}$ for the $g$th iteration as
\begin{equation}\label{eq: threshold}
{\lambda _g}({{\bm{\theta }}}) = \frac{{\sum\limits_{k=1}^{N}\sum\limits_{l=1}^{M}\omega_{kl}} - {\sum\limits_{i=1}^{g}\sum\limits_{k=1}^{N}\sum\limits_{l=1}^{M} \chi_{\mathbb{B}_{lk}({{\bm{\hat \theta }_i}})}({\bm{\theta }}) \omega_{kl}}} {\sum\limits_{k=1}^{N}\sum\limits_{l=1}^{M}\omega_{kl}}\lambda ',
%\{\lambda _g}{\rm{ = }}\frac{{{\rm{MN - }}N_g^{{\rm{elimi}}}}}{{{\rm{MN}}}}\lambda '\
\end{equation}
where $\chi_\mathbb{A}(\cdot)$ denotes the indicator function on the set $\mathbb{A}$, ${\lambda '}$ is the threshold for the original objective function ${\mathcal{F}}({{\bm{\theta }}})$ which contains all the ${MN}$ paths for ${\bm{\theta }} \in \mathbb{R}^2$ and $\omega_{kl}$ is a coefficient which accounts for the impact of the $lk$th path on the calculation of the threshold. For instance, $\omega_{kl}$ could be the intensity of noise power of the $lk$th path. If we approximately assume the coefficients are the same for all paths, then (\ref{eq: threshold}) becomes
\begin{equation}\label{eq: threshold2}
{\lambda _g}({{\bm{\theta }}}) = \frac{MN - {\sum\limits_{i=1}^{g}\sum\limits_{k=1}^{N}\sum\limits_{l=1}^{M} \chi_{\mathbb{B}_{lk}({{\bm{\hat \theta }_i}})}({\bm{\theta }}) }} {MN}\lambda',
%\{\lambda _g}{\rm{ = }}\frac{{{\rm{MN - }}N_g^{{\rm{elimi}}}}}{{{\rm{MN}}}}\lambda '\
\end{equation}
where the summation in the numerator represents the number of paths cancelled at ${\bm{\theta }}$. It means that the detection threshold can be simply computed based on the number of remaining pathes at ${\bm{\theta }}$.
%Noting that ${\lambda _g}$ is in relation to the number of paths contained in the corresponding statistical function $( {F_g}({{\bm{\hat \theta }}_g}))$, ${\lambda _g}$ is also related to  ${{\bm{\hat \theta }}_g}$. For simplicity, ${{\bm{\hat \theta }}_g}$ is given by

To summarize, this proposed SIC algorithm works in an iterative way that one target is detected and localized at one time. When a target is decided as a potential target by maximizing log-likelihood function, the objective function will be modified to clear the interference of this ``target". On the other hand, the initial detection threshold corresponding to all untreated paths can not be matched with the modified objective function composed of the remaining paths, which may result in the potential target being missed because of the higher threshold, so the detection threshold needs to be adjusted accordingly. The pseudo code of this algorithm is given in Algorithm \ref{algorithm: SIC}.
%In fact, the idea of joint detection and parameter estimation for statistical MIMO radar was mentioned before in \cite{Tajer}. However, it only dealt with a single extended target instead of multi-target. Moreover, considering the difference between the proposed SIC multi-target localization algorithm and the SIC algorithm used in communication problems such as multipath and multi-user, the former subtracts the contribution of the estimated targets from the objective unction (or the statistical function) whereas the latter cancels the interference in the received signal.
\LinesNumbered
%\linesnumbered
\begin{algorithm}[htbp]\label{algorithm: SIC}
%\dontprintsemicolon
\DontPrintSemicolon
%\SetAlgoLined
\caption{\label{alg:summary} {The Summary of SIC Algorithm}}
Compute the objective function $\mathcal{F}({\bm{\theta }})$ for the parameter space of interest ${{\bm{\theta }}} \in \mathbb{R}^2$.\;
Form the original estimate candidate set ${\Phi}_1=\mathbb{R}^2$ and the set of localized targets ${\Omega _D} = \emptyset $.
%\begin{equation}\label{eq: CC}
% {\Phi_1}= \left\{ {\bm{\theta }}: \mathcal{F}({{\bm{\theta }}}) > {\lambda ^{\prime}}, {{\bm{\theta }}} \in \mathbb{R}^2 \right\}.
% \end{equation}

\For{ $g=1,2,\ldots, G_{\max}$ }{
Obtain the $g$th maximum likelihood estimate as
${{\bm{\hat \theta }}_g} = \arg \max_{{{\bm{\theta }}} \in {\mathbb{R}^2}} \mathcal{F}_g({\bm{\theta }})$.\;
Add the $g$th estimate ${{\bm{\hat \theta }}_g}$ to the set ${\Omega _{D}}$, i.e.,${\Omega}_{D}=\{{\bm{\hat \theta }}_1,\ldots,{\bm{\hat \theta }}_g \}$.\;
Form the subset ${\Psi}({{\bm{\hat \theta }}_g})$ of candidates which share common range bins with ${{\bm{\hat \theta }}_g}$ as ${\Psi}({{\bm{\hat \theta }}_g})=\{ {\bm{\theta }}| {\bm{\theta }}\in \mathbb{B}({{\bm{\hat \theta }}_g}) \}. $\;
Update the objective function $\mathcal{F}_g({{\bm{\theta }}})$ according to the set ${\Psi}({{\bm{\hat \theta }}_g})$  by subtracting the interference of the extracted $g$th target ${{\bm{\hat \theta }}_g}$: \;
\For{ all ${{\bm{\theta }}} \in {\Phi}_1$}
{${\mathcal{F}_{g+1}}({{\bm{\theta }}}) = \mathcal{F}_g({{\bm{\theta }}}) -  {\sum\limits_{k = 1}^N {\sum\limits_{l = 1}^M {{\mathcal{M}_{lk}}({{\bm{ \theta }}_g})} } }$.\;
Recalculate the detection threshold $\lambda_g({\bm{\hat \theta }}_g)$ for ${\bm{\hat \theta }}_g$ using (\ref{eq: threshold}).\;
\If{${\mathcal{F}_{g}}({{\bm{\hat \theta }}_g})<\lambda_g({\bm{\hat \theta }}_g)$ }
{${\Omega}_{D}={\Omega}_D / {{\bm{\hat \theta }}_g}$. } } \;
\If{$g+1>G_{\max}$ }
{end the $\mathbf{for}$ loop.\;}
}
Output the set ${\Omega}_D$ containing the locations of the detected target, and the number of elements of the set ${\Omega}_D$ is the number of targets. \;
\end{algorithm}

\subsection{Discussion}
When dealing with scenarios wherein targets are \emph{completely isolated}, SSR and SIC algorithms are equivalently efficient since their required optimality assumptions are satisfied. On the other hand, with regard to the scenarios with \emph{partially separable} targets, the performance of the SSR algorithm is not guaranteed, since the local peaks of the objective function corresponding to the undetected targets may be lost because of the removal of search region. The following Proposition 1 clearly reflects the performance relationship between SSR and SIC. The proof of Proposition 1 is given in the Appendix B.

\emph{Proposition} 1: Assume a scenario with $G$ targets. Then, for vanishingly small noise, the estimation performance of SSR is upper bounded by SIC as shown in
\begin{equation}\label{eq: ssr_sic}
\sum\limits_{g = 1}^G  \mathop { \max }\limits_{ {\bm{\theta }}_g \in {\mathbb{R}^2    } } \mathcal{F}_g({\bm{\theta }}_g)  \geq
\sum\limits_{g = 1}^G  \mathop { \max }\limits_{ {\bm{\theta }}_g \in {\mathbb{S}_{g}  } } \mathcal{F}({\bm{\theta }}_g),\
\end{equation}
where the terms on the left-hand and right-hand sides of (\ref{eq: ssr_sic}) correspond to the maximum likelihood found by SIC and SSR respectively.

\begin{itemize}
  \item Essentially, the inequality in (\ref{eq: ssr_sic}) follows from the fact that the collection of all the possible sequences of estimated target locations for SSR is included in the collection of SIC. In particular, from (\ref{eq: jml_v}) and (\ref{eq: jml_v2}), we can find that the collection of all the possible sequences of the estimated $G$ target locations $[{\bm{\hat \theta }}_1, \ldots, {\bm{\hat \theta }}_G]$ for SSR and SIC are $\mathbb{S}_{1}\times\mathbb{S}_{2}\times \cdots\times{\mathbb{S}_{G} }$ and ${\mathbb{R}^{2G}}$ respectively. From (\ref{eq: S}), we have ${\mathbb{S}_{1}}=\mathbb{R}^2$, and ${\mathbb{S}_{g}}\subset\mathbb{R}^2$ for $g=2,\ldots,G$.
  \item Although SSR will generally provide inferior performance for cases with \emph{partially separable} targets, it has its own merits, i.e., simple, fast and less memory requirement. Compared to SSR, SIC must compute the modified terms and update the objective function during each iteration. Additionally, the detection threshold has to be recalculated as well.
  \item When considering the localization of moving targets, which is of considerable interest in many real-world applications, the inseperability of targets in certain transmit-receive path over a short period could be of little consequence due the change of target positions.
\end{itemize}

%In this subsection, we assume that the targets are partially separable. Thus the matrix ${\bf{\tilde S}}_{lk}^H{\bf{R}}_{lk}^{ - 1}{{\bf{\tilde S}}_{lk}}$, i.e. (\ref{eq: lh_1_1}), is not diagonal. As for the ML estimation of reflection coefficient ${{\bm{\alpha }}_{lk}}$, if the matrix ${\bf{\tilde S}}_{lk}^H{\bf{R}}_{lk}^{ - 1}{{\bf{\tilde S}}_{lk}}$ is invertible, the estimate of $\bm{\alpha}_{lk}$ can be calculated using (\ref{eq: lh_3}). We focus on such cases. Nevertheless, each one of the total $ (N_x N_y)^G $ grid points contains an inversion of a $ {{G}} \times {{G}} $ matrix (${\bf{\tilde S}}_{lk}^H{\bf{R}}_{lk}^{ - 1}{{\bf{\tilde S}}_{lk}}$), indicating a high computational complexity when many targets are present.

\section{Simulation Results}
In this section, the performances of the previously proposed SSR and SIC algorithms are investigated in three different scenarios containing both completely isolated targets and partially separable targets respectively. The following measurements are used to assess the detection and localization performance:

1) \emph{The probability of valid target detection} $P_d$: the probability that the declared target with an estimated location within $200$ m of the actual target location in both $x$ and $y$ dimensions respectively.

2) \emph{The root mean square (RMS) position error}: the average position difference between the estimated target location of the valid target and the exact location of the real target in both $x$ and $y$ dimensions respectively.

In the following analysis, the results are gathered by averaging over $1000$ Monte Carlo realizations.

\subsection{Scenario with completely isolated targets}
\begin{figure}[!t]
\centering
\includegraphics[width=6.6cm]{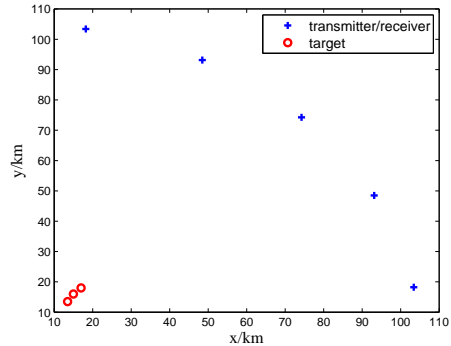}
\caption{ Sketch of the simulation scenario which contains three isolated targets and a $5\times5$ MIMO radar system, where each antenna can not only transmit, but also receive the signals from other antennas.}
\label{fig: Scenario_ci}
\end{figure}

To assess the detection and localization performance of the proposed SSR algorithm, first we consider a scenario with a $5\times5$ MIMO radar system and three completely isolated targets located at ($13.50$, $13.50$) km, ($17.00$, $18.00$) km, ($15.00$, $16.00$) km, respectively. The placement of the antennas and targets are shown in Fig. \ref{fig: Scenario_ci}, each antenna can transmit and receive a signal. The relative proportion of the square modulus of the complex amplitudes of the targets is $1:0.65:0.5$. The upper bound of the number of the potential targets is set as ${G_{\max }}=5$ in SSR.

\begin{figure}[!t]
\centering
%\subfigure[]{\includegraphics[width=6.5cm]{ndSSR.eps}}
\subfigure[]{\includegraphics[width=6.5cm]{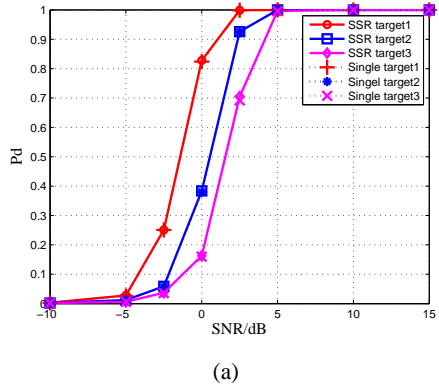}}
\caption{The detection probability $P_d$ of all targets are plotted against SNR from $-10$ dB to $15$ dB for the scenario with completely isolated targets.}
\label{SSR1}
\end{figure}

\begin{figure}[!t]
\centering
\subfigure[]{\includegraphics[width=6.5cm]{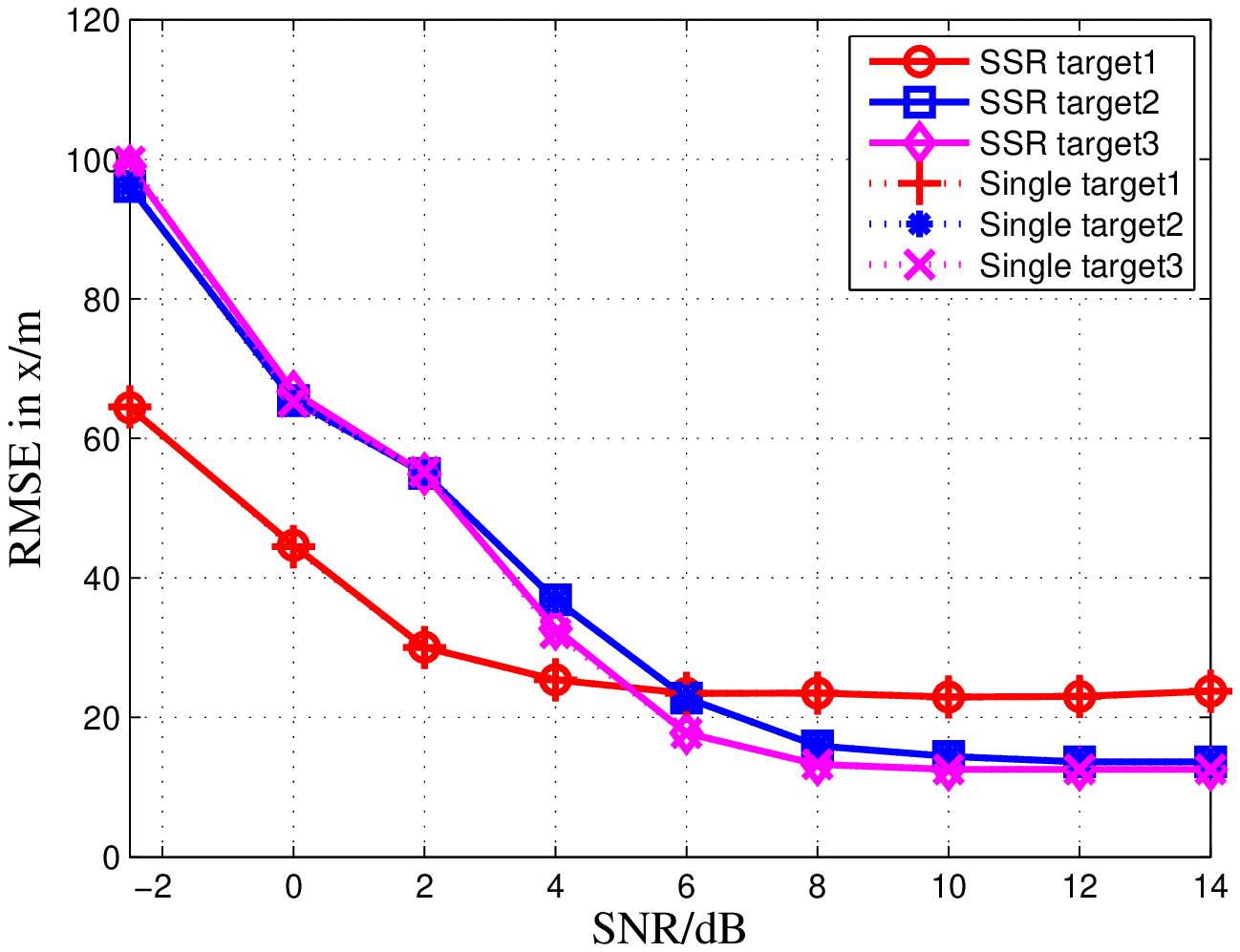}}
\subfigure[]{\includegraphics[width=6.5cm]{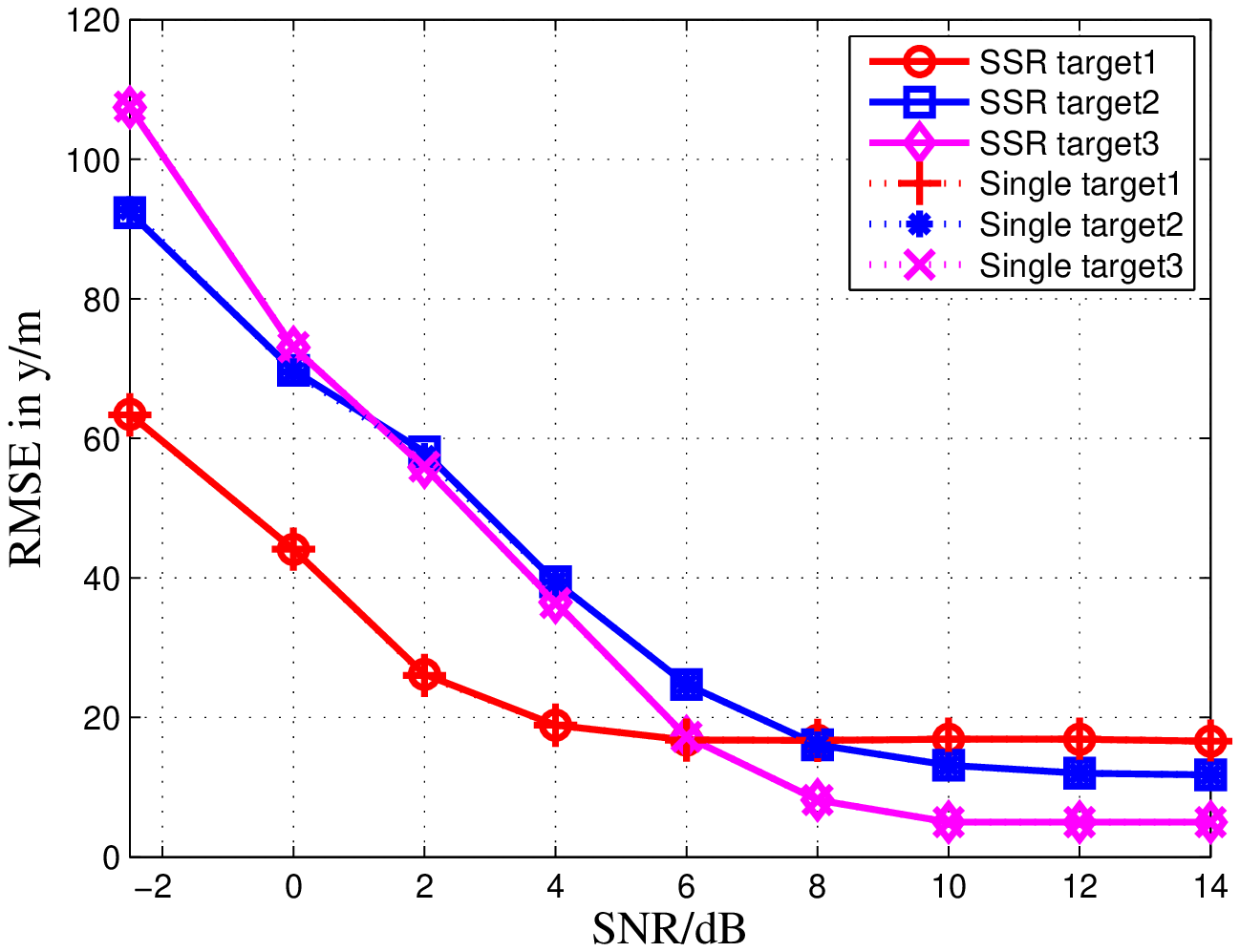}}
\caption{The RMS position errors of all targets are plotted against SNR from $-2$ dB to $14$ dB for ${P_{fa}} = {10^{-1}}$ for SSR algorithm and the scenario with completely isolated targets. (a) $x$ dimension. (b) $y$ dimension.}
\label{SSR2}
\end{figure}

The detection performance and RMS position error of the SSR algorithm is shown in Figs. \ref{SSR1} and \ref{SSR2}. To demonstrate the effectiveness of the SSR algorithm, ``the single target performances'', namely, the $P_d$ and RMS position error curves of the situation where only one specified target exists in the scenario, are also plotted to serve as a performance benchmark.

It can be seen from Fig. \ref{SSR1} that all targets can be detected with $P_d$ close to unity when the SNR exceeds $5$ dB. This shows that the SSR algorithm is able to achieve an accurate estimate for the number of targets without multi-hypothesis testing for sufficient SNR. It also indicates that the weak target (target 3) is not masked by the other strong ones. For a fixed SNR, the strong target (target 1) is more easily detected than the weak ones as expected. Moreover, the $P_d$ curves of SSR for each target are almost identical to the corresponding single target benchmark for all SNRs. This mean that the performance loss of the SSR is negligible since its required optimality assumptions are satisfied when dealing with completely isolated targets.

The localization accuracy of the SSR algorithm is shown in Fig. \ref{SSR2}, where the RMS position errors of each target for both $x$ and $y$ dimensions decrease with SNR increasing from $-2$ dB to $14$ dB. Note that the level of RMS location errors does not always follow the intensity order of the targets when SNR is high. The reason is that in the high SNR condition, the RMS location errors are very close to the Cram¡äer-Rao Bounds (CRB), which strictly depend on the geometry \cite{Godrich}. Moreover, the RMS errors change little when SNR rises from $8$ dB to $14$ dB because a grid-search method is employed in the simulation, which means that the grid width will be the main factor to restrict the localization accuracy when SNR is sufficiently large. Also, the RMS location errors curves of SSR are almost identical to the corresponding single target ones for all SNRs.

\subsection{Scenario with partially separable targets}

\begin{figure}[!t]
\centering
\includegraphics[width=7cm]{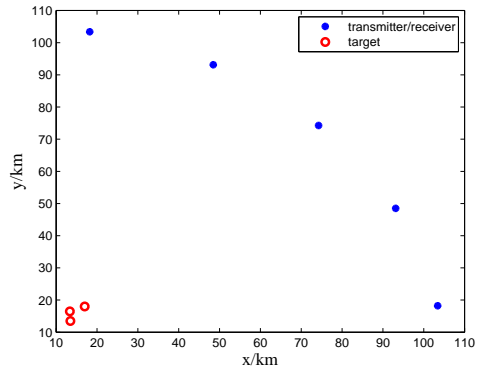}
\caption{ Sketch of the simulation scenario which contains three targets and a $5\times5$ MIMO radar system, where each antenna can not only transmit, but also receive the signals from other antennas. The positions of the three targets are carefully chosen such that two of them are inseparable in some transmit-receive paths.}
\label{fig: Scenario_pr}
\end{figure}

In this simulation, the target locations are changed to ($13.50$, $13.50$) km, ($17.00$, $18.00$) km, ($13.36$, $16.48$) km, as shown in Fig. \ref{fig: Scenario_pr}, to make sure that target 1 and target 3 are inseparable in some paths. The other parameters are set the same as before.

\begin{figure}[!t]
\centering
%\subfigure[]{\includegraphics[width=6.5cm]{fig2.eps}}
\subfigure[]{\includegraphics[width=6.5cm]{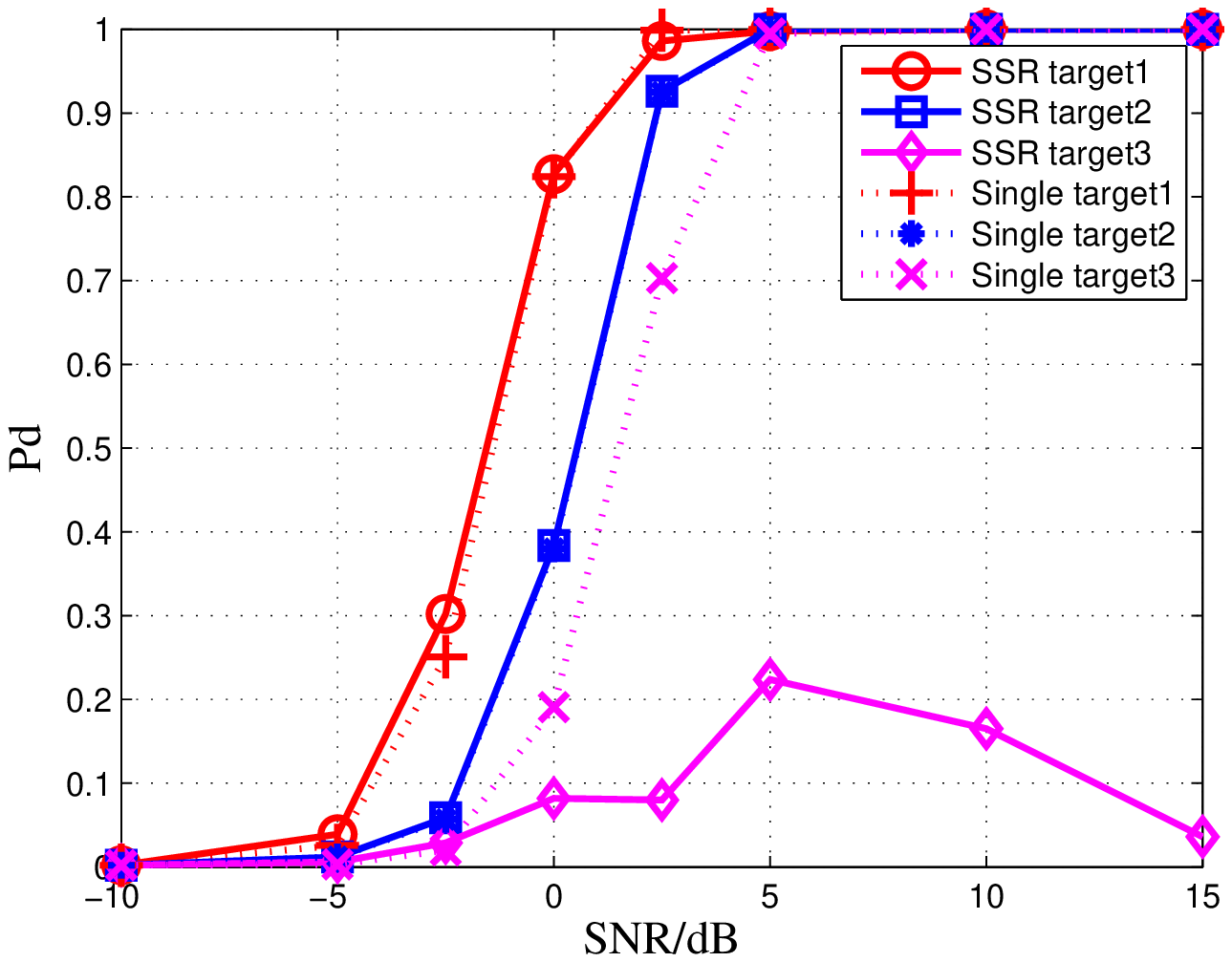}}
\subfigure[]{\includegraphics[width=6.5cm]{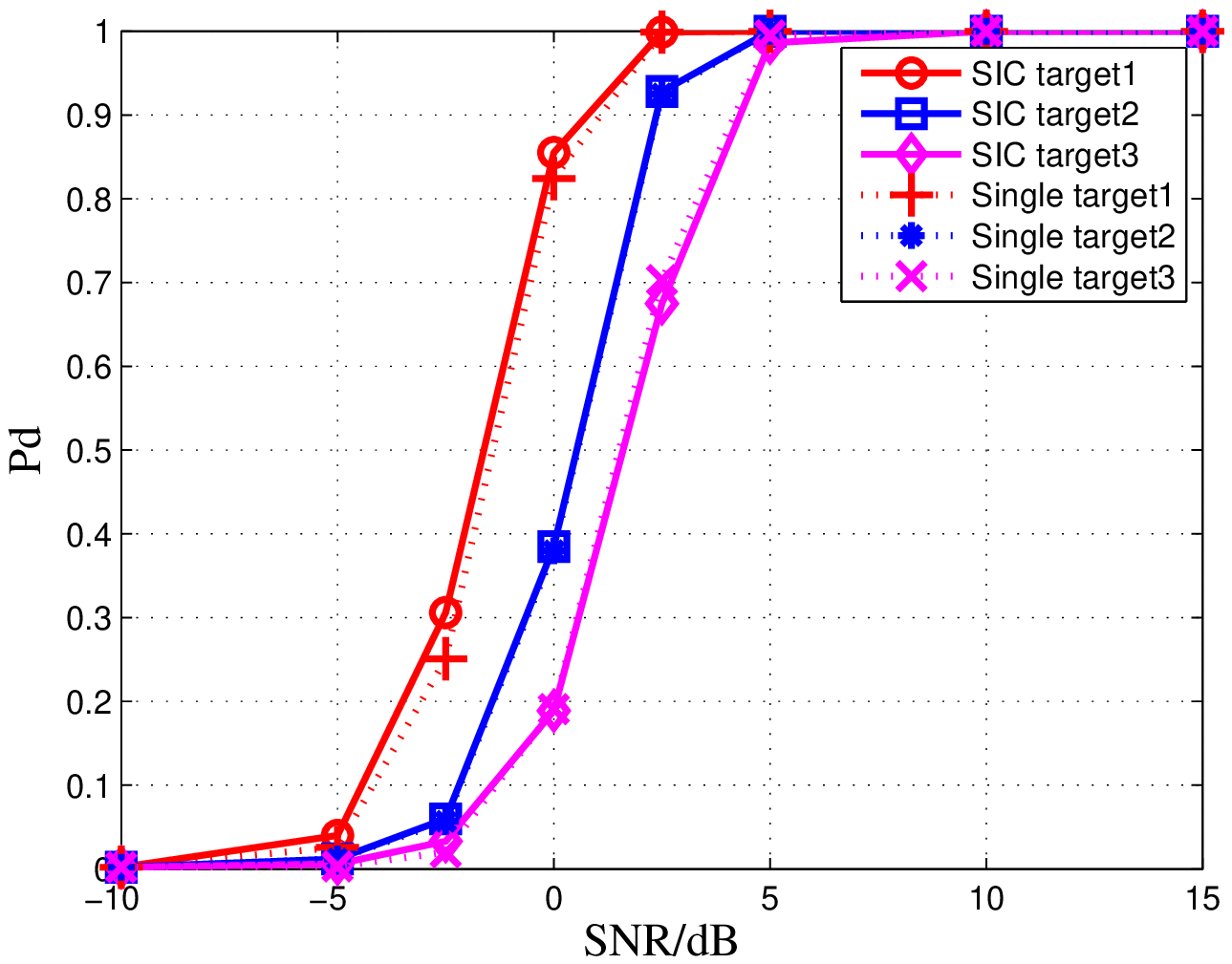}}
\caption{The detection probability $P_d$ of all targets are plotted against SNR from $-10$ dB to $15$ dB for the scenario with partially separable targets. (a) The SSR algorithm. (b) The SIC algorithm.}
\label{fig5}
\end{figure}

\begin{figure}[!t]
\centering
\subfigure[]{\includegraphics[width=6.5cm]{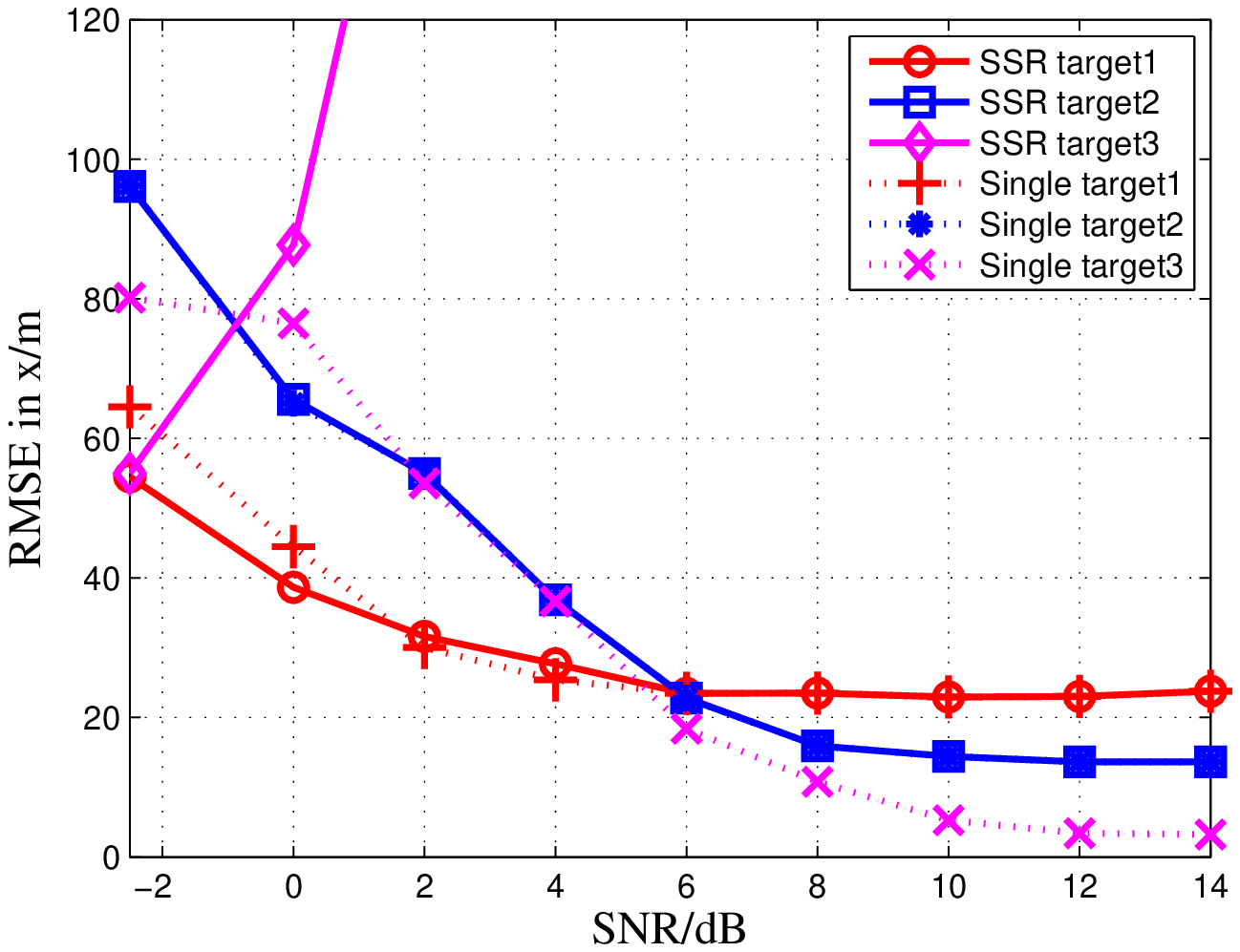}}
\subfigure[]{\includegraphics[width=6.5cm]{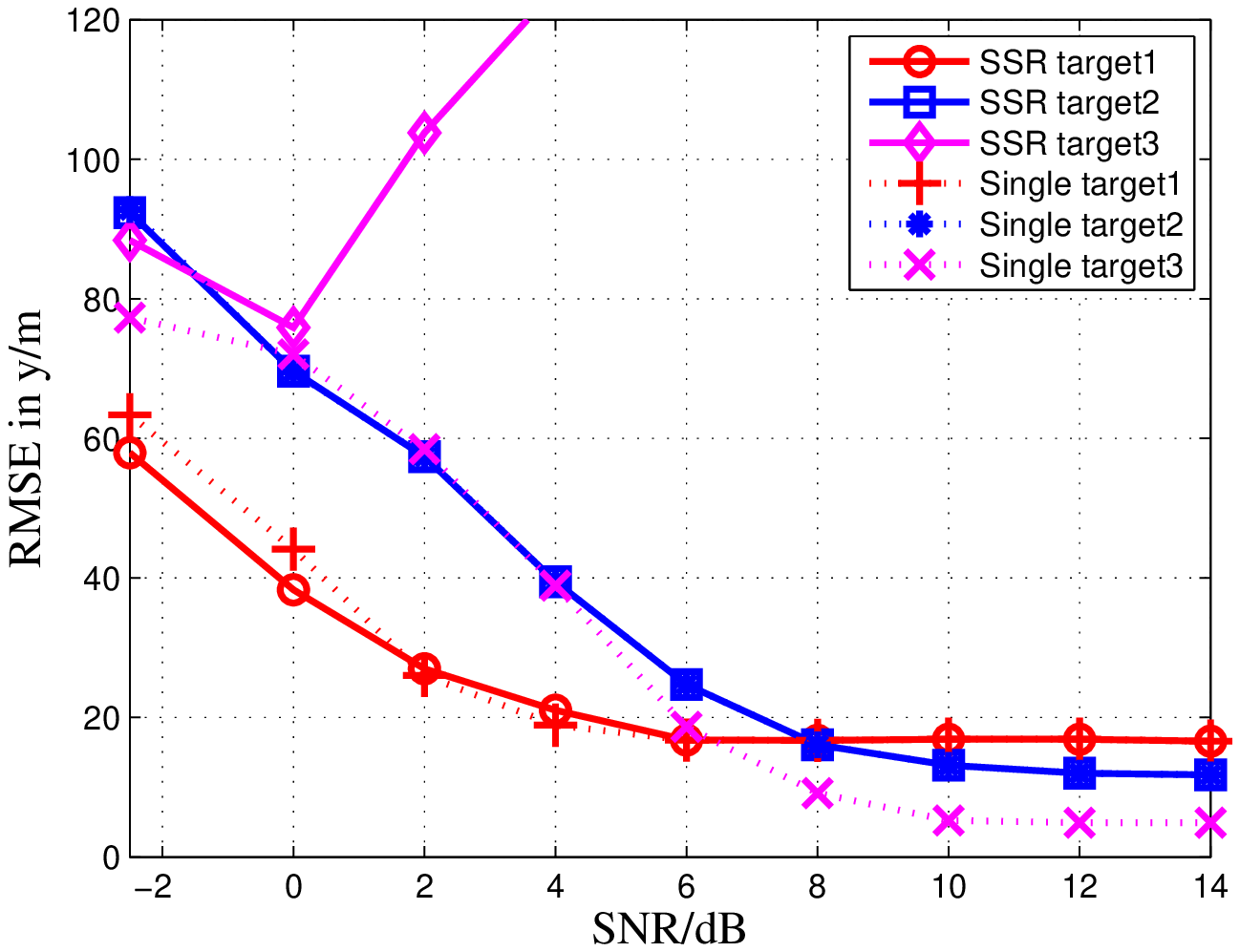}}
\caption{The RMS position errors of all targets are plotted against SNR from $-2$ dB to $14$ dB for ${P_{fa}} = {10^{ - 1}}$ with respect to SSR algorithm for the scenario with partially separable targets. (a) $x$ dimension. (b) $y$ dimension.}
\label{fig6}
\end{figure}

\begin{figure}[!t]
\centering
\subfigure[]{\includegraphics[width=6.5cm]{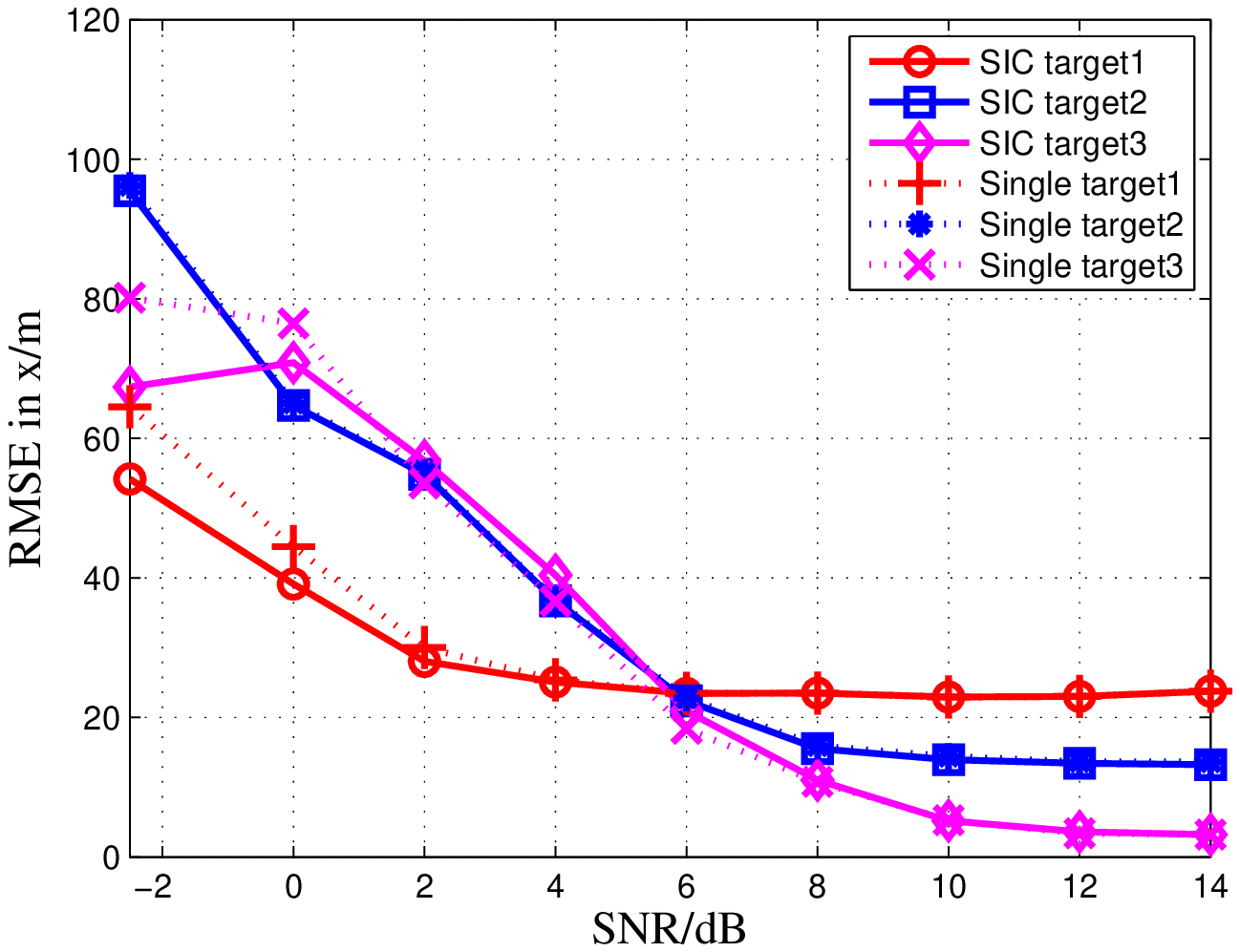}}
\subfigure[]{\includegraphics[width=6.5cm]{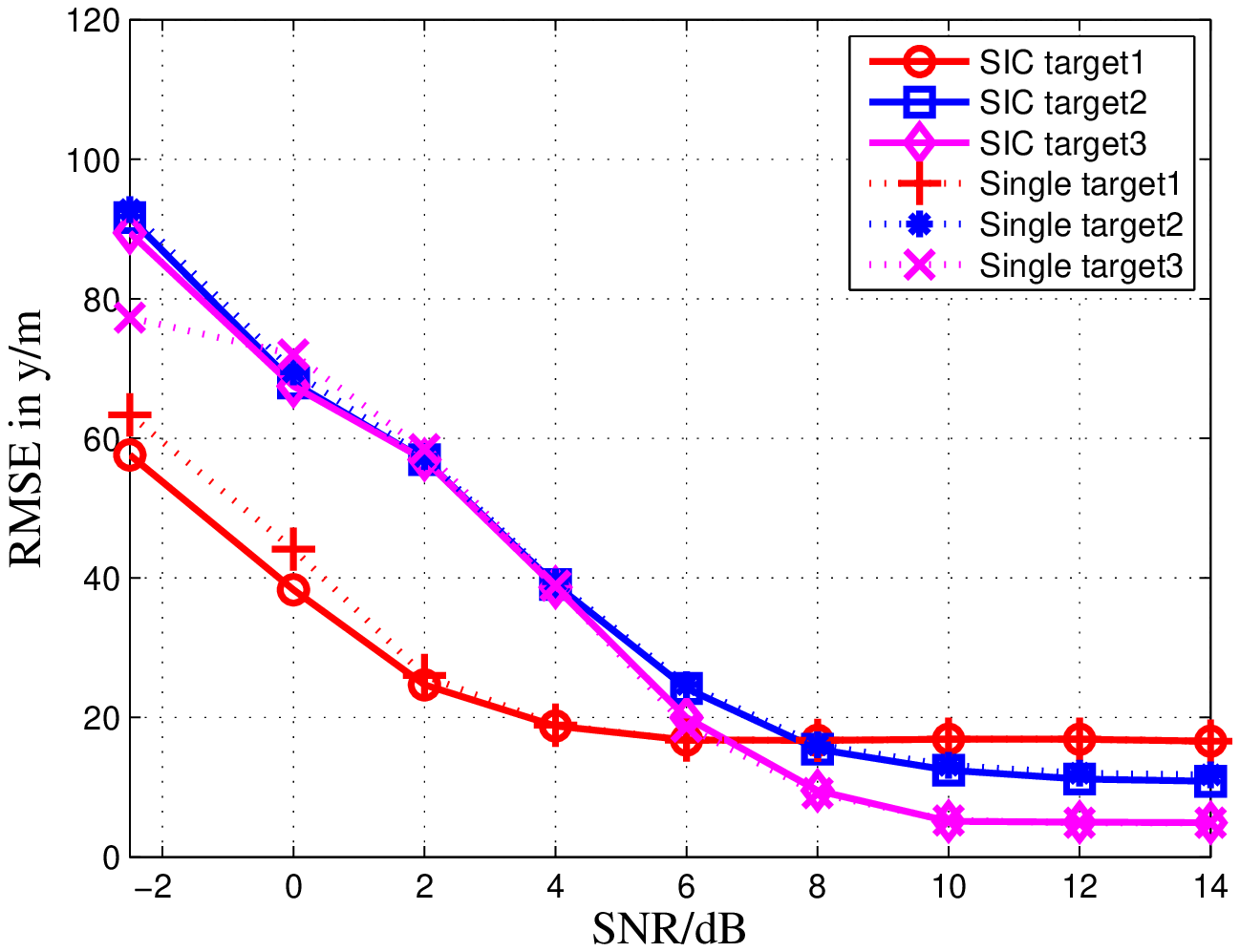}}
\caption{The RMS position errors of all targets are plotted against SNR from $-2$ dB to $14$ dB for ${P_{fa}} = {10^{ - 1}}$ for SIC algorithm and the scenario with partially separable targets. (a) $x$ dimension. (b) $y$ dimension.}
\label{fig7}
\end{figure}

The detection and localization performance of both the SSR and SIC algorithms is given in Figs. \ref{fig5}, \ref{fig6} and \ref{fig7}. In Fig. \ref{fig5}, the curves of the probability of valid detection $P_d$ are plotted against SNR from $-10$ dB to $15$ dB. As expected, SIC can deal with the partially separable targets robustly, and its $P_d$ curves of all targets are approaching unity for sufficiently high SNRs. Similar to the previous scenario, the stronger target achieves a higher detection performance than the weaker ones. By comparison, for the SSR algorithm, the detection performance of target $3$ which is the weakest one and shares common range bins with target $1$ suffers a significant performance loss due to the rude way of clearing the interference of previously detected targets.

Fig. \ref{fig6} and \ref{fig7} show the RMS position errors of all three targets for the SSR and SIC algorithms respectively. The significant performance loss of target $3$, which has overlapping paths with the first target, is clearly shown in Fig. \ref{fig6}. As opposed to the situation in Fig. \ref{fig6}, target $3$ is able to be accurately located by the SIC algorithm for a sufficient SNR. We can see that the RMS position errors curves of SIC for each targets approach the corresponding single target benchmark for almost all SNRs, indicating that SIC has the ability to accurately estimate the number of targets and localize them with quite high precision even when some targets are not isolated.


\begin{figure}[!t]
\centering
\includegraphics[width=7cm]{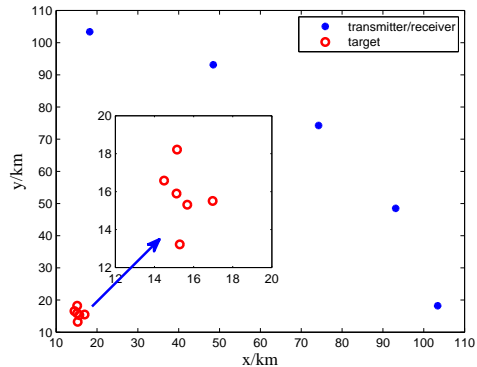}
\caption{Sketch of the simulation scenario which contains six targets and a $5\times5$ MIMO radar system, where each antenna is receiving signals transmitted from other antennas. The positions of the six targets are carefully chosen such that targets are inseparable in many transmit-receive paths.}
\label{fig: ComplexEnL}
\end{figure}

\begin{figure}[!t]
\centering
\centerline{\includegraphics[width=6.8cm]{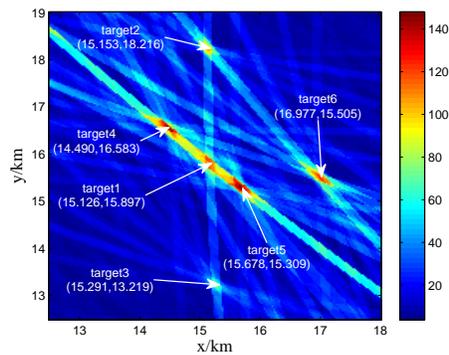}}
\caption{An illustration for the objective function for the more challenging scenario.}
\label{fig: ComplexEnLH}
\end{figure}

To assess the performance of the SIC algorithm for a more challenging scenario, this section concludes with a more complex situation involving many weak targets and overlapping paths. The number of the targets is increased to six in the scenario as shown in Fig. \ref{fig: ComplexEnL} with the position of each target given in Table I.
%which are located in  ($15.00$, $16.00$) km,($14.63$, $18.15$) km, ($15.68$, $13.32$) km,($14.46$, $16.45$) km, ($15.29$, $15.77$) km, ($12.86$, $16.97$) km,
To be more precise, targets 1, 2 and 3 share a common overlapping path, target 1 also overlaps the targets 4 and 5 in many paths, while target 6 has two common paths with target 2 and 4 respectively (see Fig. \ref{fig: ComplexEnLH} for a clearer view). The relative proportion of square modulus of the complex amplitudes of these targets is $0.5:0.5:0.5:1:1:1$. Only the case of SNR $=10$ dB is considered and $G_{\max}=6$. Fig. \ref{fig: ComplexEnLH} shows the values of the objective function in the two-dimensional plane. The RMS position errors of the SIC algorithm for all targets are shown in Table \ref{tab:table1}. The results indicate that each target can still be accurately located even though they overlap each other in many paths.
\begin{table}[!t]
\begin{center}
\caption{\label{tab:table1} The $x$ and $y$ positions (Km) of six targets for Fig. \ref{fig: ComplexEnL}}
\begin{tabular}{|c|c|c|c|c|c|c|}
  \hline
  Target & 1 & 2 & 3 & 4 & 5 & 6 \\
  \hline
  $x$ & 15.13 & 15.15 & 15.29 & 14.49 & 15.68 & 16.98 \\
  \hline
  $y$ & 15.89 & 18.21 & 13.21 & 16.58 & 15.31 & 15.51 \\
  \hline
\end{tabular}
\end{center}
\end{table}

\begin{table}[!t]
\begin{center}
\caption{\label{tab:table1} RMSE(m) of each target in $x$ and $y$ dimensions for Fig. \ref{fig: ComplexEnL} with $10$dB SNR}
\begin{tabular}{|c|c|c|c|c|c|c|}
  \hline
  Target & 1 & 2 & 3 & 4 & 5 & 6 \\
  \hline
  $x$ RMSE & 65.75 & 52.73 & 17.57 & 1.38 & 15.73 & 32.88 \\
  \hline
  $y$ RMSE & 85.71 & 31.46 & 19.02 & 1.45 & 14.37 & 36.37 \\
  \hline
\end{tabular}
\end{center}
\end{table}

\section{Conclusion}
In this paper, we consider the detection and localization of multiple targets in a noncoherent MIMO radar with widely separated antennas. To combat the troublesome high-dimensional optimization problem of simultaneously estimating multiple targets¡¯ positions, we propose two suboptimal algorithms to split the joint maximization into several disjoint optimization problems, i.e., one corresponding to each prospective target. In this way, the proposed algorithms have much lower complexity compared with the original high-dimensional estimation method. Besides, during the detection and localization process, the proposed algorithms sequentially perform single target detection after eliminating the interference in all the paths from previously declared targets, and the recursive process stops automatically if no target estimate in the current stage can exceed the detection threshold. Therefore the multi-hypothesis testing detector is no longer needed when the number of targets is unknown. Simulation results show that the proposed algorithms can correctly estimate the number of targets and localize them with high accuracy when the SNR is high. In particular, the proposed SIC algorithm works well even when some targets are not separable in some paths.

\appendices
\section{Derivation of (\ref{eq: lh_1}) and (\ref{eq: lh_2})}
The likelihood ratio function in (\ref{eq: 11}) is a scalar function in terms of the real part and the imaginary part of the complex reflection coefficient $\bm{\alpha}_{lk}$.
Respectively taking the partial derivatives for the real part $\alpha_{lkg}^{R}$ and the imaginary part $\alpha_{lkg}^{I}$ of ${\alpha}_{lkg}$(${\alpha}_{lkg}=\alpha_{lkg}^{R}+j\alpha_{lkg}^{I}$), we have
\begin{equation}\label{equ2}
\frac{\partial}{\partial\alpha_{lkg}^{R}}\ln \ell({\bf{r}}_{lk}\rm{|}\bm{\Theta},{\bm{\alpha }_{lk}})=0,
\end{equation}
\begin{equation}\label{equ3}
\frac{\partial}{\partial\alpha_{lkg}^{I}}\ln \ell({\bf{r}}_{lk} \rm{|} \bm{\Theta},{\bm{\alpha }_{lk}})=0.
\end{equation}\
By substituting (\ref{eq: 11}) into (\ref{equ2}) we have,
\begin{equation}\label{equ4}
\begin{split}
\frac{1}{2}{\bf{r}}_{lk}^{H}{\bf{R}}_{lk}^{ - 1}{\bf{\tilde{s}}}_{lkg}&+\frac{1}{2}{\bf{\tilde{s}}}_{lkg}^{H}{\bf{R}}_{lk}^{ - 1}{\bf{r}}_{lk}\\
& -\frac{1}{2}\left[{\bf{\tilde{s}}}_{lkg}^{H}{\bf{R}}_{lk}^{ - 1}\left(\sum_{g=1}^{G}\alpha_{lkg}{\bf{\tilde{s}}}_{lkg}\right)\right.\\
& +\left.\left(\sum_{g=1}^{G}\alpha_{lkg}{\bf{\tilde{s}}}_{lkg}\right)^{H}{\bf{R}}_{lk}^{ - 1}{\bf{\tilde{s}}}_{lkg}\right]=0.
\end{split}
\end{equation}
Then by isolating the term related to the complex reflection coefficient of the $g$th target from the summation terms in (\ref{equ4}), we have,
\begin{equation}\label{equ5}
\begin{split}
\frac{1}{2}{\bf{r}}_{lk}^{H}{\bf{R}}_{lk}^{ - 1}{\bf{\tilde{s}}}_{lkg}&+\frac{1}{2}{\bf{\tilde{s}}}_{lkg}^{H}{\bf{R}}_{lk}^{ - 1}{\bf{r}}_{lk}\\
&-\frac{1}{2}\left[{\bf{\tilde{s}}}_{lkg}^{H}{\bf{R}}_{lk}^{ - 1} \left( \sum_{g_1=1,g_1\neq{g}}^{G}\alpha_{lkg_1}{\bf{\tilde{s}}}_{lkg_1} \right)\right.\\
&+{\bf{\tilde{s}}}_{lkg}^{H}{\bf{R}}_{lk}^{ - 1}\alpha_{lkg}{\bf{\tilde{s}}}_{lkg}\\
&+\left(\sum_{g_1=1,g_1\neq{g}}^{G}\alpha_{lkg_1}{\bf{\tilde{s}}}_{lkg_1}\right)^{H}{\bf{R}}_{lk}^{ - 1}{\bf{\tilde{s}}}_{lkg}\\
&+\left.\alpha_{lkg}^{\ast}{\bf{\tilde{s}}}_{lkg}^{H}{\bf{R}}_{lk}^{ - 1}{\bf{\tilde{s}}}_{lkg}\right]=0.
\end{split}
\end{equation}
Further by combining the terms in (\ref{equ5}) as below,
\begin{equation}\label{equ6}
\alpha_{lkg}{\bf{\tilde{s}}}_{lkg}^{H}{\bf{R}}_{lk}^{ - 1}{\bf{\tilde{s}}}_{lkg} + \alpha_{lkg}^{\ast}{\bf{\tilde{s}}}_{lkg}^{H}{\bf{R}}_{lk}^{ - 1}{\bf{\tilde{s}}}_{lkg}=2\alpha_{lkg}^{R}{\bf{\tilde{s}}}_{lkg}^{H}{\bf{R}}_{lk}^{ - 1}{\bf{\tilde{s}}}_{lkg},
\end{equation}
one can simplified (\ref{equ5}) as,
\begin{equation}\label{equ7}
\begin{split}
\frac{1}{2}{\bf{r}}_{lk}^{H}{\bf{R}}_{lk}^{ - 1}{\bf{\tilde{s}}}_{lkg} &+ \frac{1}{2}{\bf{\tilde{s}}}_{lkg}^{H}{\bf{R}}_{lk}^{ - 1}{\bf{r}}_{lk}\\
&-\frac{1}{2}\left[{\bf{\tilde{s}}}_{lkg}^{H}{\bf{R}}_{lk}^{ - 1}\left(\sum_{g_1=1,g_1\neq{g}}^{G}\alpha_{lkg_1}{\bf{\tilde{s}}}_{lkg_1}\right)\right.\\
&+2\alpha_{lkg}^{R} {\bf{\tilde{s}}}_{lkg}^{H}{\bf{R}}_{lk}^{ - 1} {\bf{\tilde{s}}}_{lkg} \\
& +\left. \left(\sum_{g_1=1,g_1\neq{g}}^{G}\alpha_{lkg_1}{\bf{\tilde{s}}}_{lkg_1}\right)^{H}{\bf{R}}_{lk}^{ - 1}{\bf{\tilde{s}}}_{lkg}\right]=0.
\end{split}
\end{equation}
Similarly, the partial derivative for the imaginary part $\alpha_{lkg}^{I}$ of the complex reflection coefficient, namely (\ref{equ3}), has the following expression,
\begin{equation}\label{equ8}
\begin{split}
\frac{j}{2}{\bf{r}}_{lk}^{H}{\bf{R}}_{lk}^{ - 1}{\bf{\tilde{s}}}_{lkg} &- \frac{j}{2}{\bf{\tilde{s}}}_{lkg}^{H}{\bf{R}}_{lk}^{ - 1}{\bf{r}}_{lk}\\
&-\frac{1}{2}\left[{-j}{\bf{\tilde{s}}}_{lkg}^{H}{\bf{R}}_{lk}^{ - 1}\left(\sum_{g_1=1,g_1\neq{g}}^{G}\alpha_{lkg_1}{\bf{\tilde{s}}}_{lkg_1}\right)\right. \\ &+2\alpha_{lkg}^{I} {\bf{\tilde{s}}}_{lkg}^{H}{\bf{R}}_{lk}^{ - 1}{\bf{\tilde{s}}}_{lkg}\\
& +\left. {j}\left(\sum_{g_1=1,g_1\neq{g}}^{G}\alpha_{lkg_1}{\bf{\tilde{s}}}_{lkg_1}\right)^{H}{\bf{R}}_{lk}^{ - 1}{\bf{\tilde{s}}}_{lkg}\right]=0.
\end{split}
\end{equation}
Combining (\ref{equ7}) and (\ref{equ8}), we have, after some working,
\begin{equation}\label{equ10}
\sum\limits_{{g}_{1} = 1}^G {\alpha _{lk{g}_{1}}}{{\bf{\tilde s}}_{lkg}^H{\bf{R}}_{lk}^{ - 1}}{{{\bf{\tilde s}}}_{lk{g}_{1}}} = {\bf{\tilde s}}_{lkg}^H{\bf{R}}_{lk}^{ - 1}{\bf{r}}_{lk}.
\end{equation}
Thus (\ref{eq: lh_1}) is proofed. Also one can find that (\ref{equ10}) is the $g$th of the $G$ equations constructing (\ref{eq: lh_2}). Combination of the $G$ equations into matrix formation using term ${{\bf{\tilde S}}_{lk}} = [{{\bf{\tilde s}}_{lk1}},{{\bf{\tilde s}}_{lk2}}, \ldots, {{\bf{\tilde s}}_{lkG}}]$ yields (\ref{eq: lh_2}) in this paper.

\section{Proof of Proposition $1$}
Let ${{\bm{\theta}}_g^T}$, $g=1, \ldots, G$ denotes the true position of the $g$th target. For vanishingly small noise, the value of target related objective functions are far greater than noise related ones. Thus, when considering the scenario with isolated targets, all the targets can be localized one by one since the previously detected targets will not affect the subsequent detection and localization of remaining targets, namely, ${{\bm{\theta}}_{i+1}^T}$, $\ldots$, ${{\bm{\theta}}_G^T} \notin {\mathbb{B}({{\bm{\theta }_1}})} \bigcup \ldots \bigcup{\mathbb{B}({{\bm{\theta }_i}})}$ for $i=1,\ldots G-1$. Therefore we have,
\begin{equation}
\sum\limits_{g = 1}^G  \mathop { \max }\limits_{ {\bm{\theta }}_g \in {\mathbb{R}^2    } } \mathcal{F}_g({\bm{\theta }}_g)  =
\sum\limits_{g = 1}^G  \mathop { \max }\limits_{ {\bm{\theta }}_g \in {\mathbb{S}_{g}  } } \mathcal{F}({\bm{\theta }}_g)\ =
\sum\limits_{g = 1}^G  \mathop  \mathcal{F}({\bm{\theta }}_g^T)
\end{equation}

With regard to the scenario with partially separable targets, i.e., targets $A$ and $B$ are inseparable in one or more paths. Once one of the two targets has been localized (say target $A$, without loss of generality), then the true position of target $B$ is eliminated from the search space for SSR, while SIC only eliminates the interference of the inseparable paths. Suppose that target $B$ is found by SIC at the $i$th iteration with objective function $\mathcal{F}_i({\bm{\theta }}^B)=\sum\limits_{d = 1}^D  { {\ell_{d}({{\bm{\theta }^B}})} }$, where ${\ell_{d}({{\bm{\theta }^B}})}$ is the log-likelihood function of the $d$th transmit-receive path. Note that the number $D$ of the remaining log-likelihood functions for ${\bm{\theta }^B}$ is less than $MN$ due the update of the objective function (\ref{eq: m_1}). Term $\mathcal{F}_i({\bm{\theta }}^B)$ can be viewed as a positive contribution to the summation on the left-hand side of (\ref{eq: ssr_sic}). However, the localization of target $B$ can also result negative impact to the summation of the objective function of SIC if ${{\mathbb{B}}({{\bm{ \theta }}^B})}$ covers any undetected targets. The negative impact by keeping target $B$ can be expressed as  $\sum\limits_{d = 1}^D  { {\ell_{d}({{\bm{\theta }^B}})} r_d }$, where $r_d$ denotes the number of targets covered by ${{\mathbb{B}}_d({{\bm{ \theta }}^B})}$ and $\sum\limits_{d = 1}^D{r_d \leq G-i}$. It can be found that the positive impact by keeping target $B$ is always great than or equal to its negative impact.

In summary, combine the two cases above, inequality (\ref{eq: ssr_sic}) is proved.

\end{document}